\documentclass[
superscriptaddress,
aps,
pra,
twocolumn
]{revtex4-1}

\usepackage{bm}% bold math
\usepackage{ulem}
\usepackage{epsfig}
\usepackage{graphicx}% Include figure files
\usepackage{amssymb,amsmath,amsbsy,amsgen,amsfonts}    
\usepackage{dcolumn}% Align table columns on decimal point
\usepackage{amsthm}
\usepackage{mathrsfs}
\usepackage{latexsym}
\usepackage{array}
\usepackage{color}
\usepackage{amstext}
\allowdisplaybreaks[1]
\usepackage{txfonts}
\usepackage{pifont}
\usepackage{braket}
\usepackage{epstopdf}
\usepackage{setspace} %%besserer marginpar: \mnote

\newcommand{\abs}[1]{\left\vert #1\right\vert}

\DeclareSymbolFont{symbols}{OMS}{cmsy}{m}{n}

\begin{document}
%\preprint{APS/123-QED}
\title{Optimal Gaussian measurements for phase estimation in single-mode Gaussian metrology}
%\author{Changhun Oh}%
\author{C. Oh}%
%\email{v55ohv@gmail.com}
\affiliation{Center for Macroscopic Quantum Control, Department of Physics and Astronomy, Seoul National University, Seoul 08826, Korea}

%\author{Changhyoup Lee}
\author{C. Lee}
\email{changhyoup.lee@gmail.com}
\affiliation{Institute of Theoretical Solid State Physics, Karlsruhe Institute of Technology, 76131 Karlsruhe, Germany}

%\author{Carsten Rockstuhl}
\author{C. Rockstuhl}
\affiliation{Institute of Theoretical Solid State Physics, Karlsruhe Institute of Technology, 76131 Karlsruhe, Germany}
\affiliation{Institute of Nanotechnology, Karlsruhe Institute of Technology, 76021 Karlsruhe, Germany}

%\author{\\Hyunseok Jeong}
\author{H. Jeong}%
\affiliation{Center for Macroscopic Quantum Control, Department of Physics and Astronomy, Seoul National University, Seoul 08826, Korea}

%\author{Jaewan Kim}
\author{J. Kim}
\affiliation{School of Computational Sciences, Korea Institute for Advanced Study, Hoegi-ro 85, Dongdaemun-gu, Seoul 02455, Korea}

%\author{Hyunchul Nah}
\author{H. Nha}
\affiliation{School of Computational Sciences, Korea Institute for Advanced Study, Hoegi-ro 85, Dongdaemun-gu, Seoul 02455, Korea}
\affiliation{Department of Physics, Texas $A\&M$ University at Qatar, Education City, P.O.Box 23874, Doha, Qatar}

%\author{Su-Yong Lee}%
\author{S.-Y. Lee}%
\email{papercrane79@gmail.com}
\affiliation{School of Computational Sciences, Korea Institute for Advanced Study, Hoegi-ro 85, Dongdaemun-gu, Seoul 02455, Korea}

\date{\today}% It is always \today, today,
% but any date may be explicitly specified

\begin{abstract} 
The central issue in quantum parameter estimation is to find out the optimal measurement setup that leads to the ultimate lower bound of an estimation error. We address here a question of whether a Gaussian measurement scheme can achieve the ultimate bound for phase estimation in single-mode Gaussian metrology that exploits single-mode Gaussian probe states in a Gaussian environment. We identify three types of optimal Gaussian measurement setups yielding the maximal Fisher information depending on displacement, squeezing, and thermalization of the probe state. We show that the homodyne measurement attains the ultimate bound for both displaced thermal probe states and squeezed vacuum probe states, whereas for the other single-mode Gaussian probe states, the optimized Gaussian measurement cannot be the optimal setup, although they are sometimes nearly optimal. We then demonstrate that the measurement on the basis of the product quadrature operators $\hat{X}\hat{P}+\hat{P}\hat{X}$, i.e., a non-Gaussian measurement, is required to be fully optimal. 
\end{abstract}


\maketitle
%\tableofcontents

%%%%%%%%%%%%%%%%%%%%%%%%%%%%%%%%%%%%%%%%%%%%%%%%%%%%%%%%%%%%%%%%%%%%%%%%
%%%%%%%%%%%%%%%%%%%%%%%%%%%%%%%%%%%%%%%%%%%%%%%%%%%%%%%%%%%%%%%%%%%%%%%%
\section{Introduction}
Gaussian states are useful resources in quantum optical technology~\cite{Ferraro05, Wang07, Weedbrook12, Adesso14}. Their intrinsic features that enable full analytical calculations for any Gaussian states and operations have attracted intensive interests from the theoretical perspective in many scientific areas. Furthermore, their experimental control is less demanding compared to those required for non-Gaussian states such as Fock states. Consequently, they offer much promising building blocks for quantum information processing from a practical point of view. Such fascinating aspects have boosted both theoretical and experimental studies with Gaussian states over the last decade in a broad range from fundamentals to applications. 

Gaussian states are often cooperated with the so-called Gaussian measurements, defined as a measurement scheme that produces a Gaussian probability distribution of outcomes for any Gaussian state~\cite{Weedbrook12}. Typical Gaussian measurements are the homodyne and heterodyne measurements, but a general Gaussian positive-operator-valued measure (POVM) can also be constructed~\cite{Braunstein05}. Gaussian measurements enable the full characterization of all Gaussian states~\cite{Lvovsky09}, so that they can be used for testing a necessary and sufficient condition for the inseparability of Gaussian states~\cite{Duan00,Simon00}. It has been demonstrated that Gaussian measurements sufficiently constitute the optimal set of POVMs for a minimization involved in the computation of quantum discord for Gaussian states~\cite{Giorda12, Pirandola14}. In particular, the homodyne detection has offered not only an optimal tool to distinguish two pure single-mode Gaussian states~\cite{Nha05}, but also a nearly optimal estimation of Gaussian quantum discord for small values of discord~\cite{Blandino12}. On the other hand, it has also been shown that Gaussian states cannot be distilled by local Gaussian operations with classical communications~\cite{Eisert02, Fiurasek02}. Moreover, the violation of the Bell inequality requires non-Gaussian measurements~\cite{Nha04,Garcia-Patron04}, and there also exist two-mode Gaussian states whose quantum steering can be demonstrated only by non-Gaussian measurements~\cite{Ji16, Wollman16}. Thus, a question arises in the context of quantum metrology: Are Gaussian measurements a sufficient tool for Gaussian metrology, where the parameter being estimated is encoded to Gaussian probe states?
%~\cite{Giovannetti04, Boto00, Giovannetti06, Giovannetti11}.

\begin{figure}[b]
\centering
\includegraphics[width=0.45\textwidth]{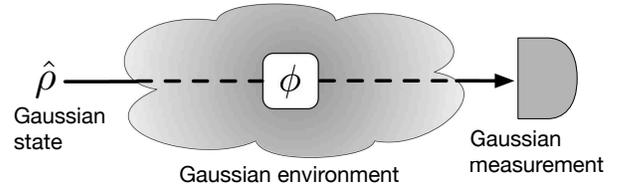}
\caption{Scheme of a fully Gaussian single-mode metrology. The Gaussian probe state evolves under Gaussian environment, where the parameter $\phi$ being estimated is encoded to the probe state. The output state is then analyzed by a Gaussian measurement.}
\label{setup}
\end{figure}

In this work, we address this question by considering a fully Gaussian single-mode metrology for phase estimation, as depicted in Fig.~1. To our aim, an arbitrary single-mode Gaussian probe state is considered to undergo a phase operation in a Gaussian noise environment. The phase-shifted probe state is then analyzed by Gaussian measurements, characterized by control parameters being optimized in order to minimize the estimation error, or equivalently to maximize the associated Fisher information~(FI)~\cite{Braunstein94,Braunstein96}. The maximal FI obtained by the optimized Gaussian measurement sets the minimum bound of the estimation error according to the Cram\'er-Rao inequality. We compare such minimum bounds with the ultimate bound calculated by quantum Fisher information (QFI), the FI maximized over all POVMs, including non-Gaussian measurements~\cite{Braunstein94, Braunstein96}. As a result, we find that there exist three types of optimal Gaussian measurements depending on displacement, squeezing and thermalization of the probe state. We also show that the optimally chosen Gaussian measurements enable to achieve the ultimate error bound when a phase information is encoded in a displaced thermal state, or squeezed vacuum state, while non-Gaussian measurements are required for the other kinds of single-mode Gaussian states to attain the ultimate bound. We then prove that the required non-Gaussian measurement is the POVMs constructed over the eigenbasis of the product quadrature operators $\hat{X}\hat{P}+\hat{P}\hat{X}$. The results of this work not only cover all partial results that have been discussed so far in the literature~\cite{monras2006,olivares2009,Monras2013,Pinel13,Jiang14,Matsubara18} (as shall be explained in detail throughout this work), but also offer rich conclusive discussions, with the full generality, regarding phase estimation using single-mode Gaussian states. We thus expect our general study to be fundamentally interesting, and also practically useful in cases where metrological resources are limited.

%The paper is structured as follows. In Sec.~II, we introduce the metrological scenario that uses a single-mode Gaussian state and Gaussian measurement under Gaussian environment. In Sec.~III, both classical and quantum Cram\'er-Rao bounds are introduced to be used for quantifying the estimation error of our phase estimation. In Sec.~IV, we investigate the estimation performance with Gaussian measurements for a displaced thermal state, a squeezed thermal state, and a displaced squeezed thermal state. Explicit forms of optimized Gaussian measurements for the respective cases are shown and explained in detail. In Sec.~V, we discuss why the optimized Gaussian measurement cannot be the optimal setup for particular types of Gaussian states. Finally, in Sec.~VI, we summarize our work and conclude with an outlook on future studies. 

%%%%%%%%%%%%%%%%%%%%%%%%%%%%%%%%%%%%%%%%%%%%%%%%%%%%%%%%%%%%%%%%%%%%%%%%
%%%%%%%%%%%%%%%%%%%%%%%%%%%%%%%%%%%%%%%%%%%%%%%%%%%%%%%%%%%%%%%%%%%%%%%%
%\subsection{Single-mode Gaussian metrology}
%%%%%%%%%%%%%%%%%%%%%%%%%%%%%%%%%%%%%%%%%%%%%%%%%%%%%%%%%%%%%%%%%%%%%%%%
%%%%%%%%%%%%%%%%%%%%%%%%%%%%%%%%%%%%%%%%%%%%%%%%%%%%%%%%%%%%%%%%%%%%%%%%
In a fully Gaussian single-mode metrology for parameter estimation as depicted in Fig.~1, we employ a single-mode Gaussian state as a probe state, and let it evolve under the influence of a Gaussian environment in which the encoding of a parameter also takes place. We aim to estimate the parameter $\phi$ while minimizing the associated estimation error by choosing an optimal Gaussian measurement setup. The parameter $\phi$ being encoded can be an optical phase, loss rate, squeezing parameter, temperature, frequency, and so on. In this work, we particularly choose a single-mode phase for the parameter $\phi$ while leaving the studies on the other types of parameter estimation for future works. In the following we shortly discuss the different ingredients to be considered.

%%%%%%%%%%%%%%%%%%%%%%%%
%\subsection{Gaussian probe state}
%%%%%%%%%%%%%%%%%%%%%%%%
%\textit{Gaussian probe state ---.} 
Any single-mode Gaussian state can always be written by a displaced squeezed thermal state~\cite{Ferraro05, Wang07, Weedbrook12}, defined as
\begin{align}
\hat{\rho}_{\rm in}&=\hat{D}(\alpha_{\rm in})\hat{S}(\xi_{\rm in})\hat{\rho}_{T}(n_{\rm th,in})\hat{S}^\dagger(\xi_{\rm in})\hat{D}^\dagger(\alpha_{\rm in}),
\label{input}
%&=\sum_{n=0}^\infty p_n\hat{D}(\alpha)\hat{S}(\xi)|n\rangle\langle n|\hat{S}^\dagger(\xi)\hat{D}^\dagger(\alpha) \label{spectral}
\end{align}
where 
%$\hat{\rho}_T=\exp(-y\hbar\omega\hat{a}^\dagger\hat{a})/\text{Tr}[\exp(-y\hbar\omega\hat{a}^\dagger\hat{a})]=\sum_{n=0}^\infty p_n|n\rangle\langle n|$ 
$\hat{\rho}_{T}(n_{\rm th,in})$ denotes a thermal state with an average photon number of $n_\text{th,in}={\rm Tr}[\hat{n}\hat{\rho}_{T}(n_{\rm th,in})]$,
$\hat{D}(\alpha_{\rm in})=\exp(\alpha_{\rm in} \hat{a}^\dagger-\alpha_{\rm in}^* \hat{a})$ is a displacement operator with $\alpha_{\rm in}=\abs{\alpha_{\rm in}}e^{i\theta_{\rm c}}$, and $\hat{S}(\xi_{\rm in})=\exp(\frac{1}{2}\xi_{\rm in}^*\hat{a}^2 -\frac{1}{2}\xi_{\rm in} \hat{a}^{\dagger2})$ is a squeezing operator with $\xi_{\rm in}=r_{\rm in} e^{i\theta_{\rm s}}$ for $r_{\rm in}\ge0$. 
%$n_\text{th}=\langle\hat{n}\rangle=(\exp(y\hbar\omega)-1)^{-1}$
% $p_n=n_\text{th}^n/(1+n_\text{th})^{n+1}$.
%Note that Eq.~(\ref{spectral}) is a spectral decomposition of an arbitrary Gaussian state since $\hat{D}(\alpha)\hat{S}(\xi)|n\rangle$ and $\hat{D}(\alpha)\hat{S}(\xi)|m\rangle$ are orthonormal to each other when $n\neq m$. 
A Gaussian state is known to be characterized in terms of, by definition, only the first and second moments. So it is often convenient to rewrite a single-mode Gaussian state of Eq.~\eqref{input} by the covariance matrix $\boldsymbol{\sigma}$ and the displacement vector $\boldsymbol{d}$, defined as $\sigma_{jk}=\langle\{\hat{x}_{j}-\langle\hat{x}_{j}\rangle,\hat{x}_{k}-\langle\hat{x}_{k}\rangle\}\rangle/2$, and $d_{j}=\langle \hat{x}_{j}\rangle$, respectively, for the quadrature operators $\hat{x}_{1}=(\hat{a}+\hat{a}^{\dagger})/\sqrt{2}$ and $\hat{x}_{2}=(\hat{a}-\hat{a}^{\dagger})/\sqrt{2}i$. The latters also read as $\hat{x}_{1}=\hat{X}_{0}$ and $\hat{x}_{2}=\hat{P}_{0}$~(or~$\hat{X}_{\pi/2}$), where the rotated quadrature operator is given by $\hat{X}_{\theta}=\hat{R}^{\dagger}(\theta)\hat{X}\hat{R}(\theta)$~[or~$\hat{P}_{\theta}=\hat{R}^{\dagger}(\theta)\hat{P}\hat{R}(\theta)$] and $\hat{R}(\theta)=e^{-i\theta\hat{a}^{\dagger}\hat{a}}$. The $\boldsymbol{\sigma}_{\rm in}$ and $\boldsymbol{d}_{\rm in}$ for the input state of Eq.~\eqref{input} read as
\begin{align}
\boldsymbol{\sigma}_{\rm in}
&=\frac{2n_\text{th,in}+1}{2}\nonumber\\
&~\times\begin{pmatrix}
\cosh{2r_{\rm in}}-\sinh{2r_{\rm in}}\cos\theta_{\rm s} & -\sinh{2r_{\rm in}}\sin\theta_{\rm s}\\
-\sinh{2r_{\rm in}}\sin\theta_{\rm s} & \cosh{2r_{\rm in}}+\sinh{2r_{\rm in}}\cos\theta_{\rm s}
\end{pmatrix}\label{input_sigma}, \\
\boldsymbol{d}_{\rm in}&=
\sqrt{2}~
\begin{pmatrix}
\abs{\alpha_{\rm in}}\cos\theta_{\rm c} \\
\abs{\alpha_{\rm in}}\sin\theta_{\rm c}
\end{pmatrix}\label{input_d}.
\end{align}
The average number of photons in a single-mode Gaussian state of Eq.~\eqref{input} is then written as
%\begin{align*}
$N=\frac{1}{2}\left({\rm Tr}\left[\boldsymbol{\sigma}\right]+\left\vert\boldsymbol{d}\right\vert^{2}-1\right)$.
%&=\frac{1}{2}\left[ (2n_{\rm th}+1)\cosh2r+2\alpha^{2}-1 \right].
%\end{align*}

%%%%%%%%%%%%%%%%%%%%%%%%
%\subsection{Phase shift}
%%%%%%%%%%%%%%%%%%%%%%%%
%\textit{Phase shift ---.}
We suppose that a phase shift by an operator $\hat{R}(\phi)$ occurs to the Gaussian probe state of Eq.~\eqref{input}. The phase shifter transforms the covariance matrix and displacement vector in a way that $\theta_{\rm s}\rightarrow \theta_{\rm s}-2\phi$ 
in Eq.~\eqref{input_sigma} and $\theta_{\rm c}\rightarrow \theta_{\rm c}-\phi$
%\begin{align*}
%\theta_{\rm s}&\rightarrow \theta_{\rm s}-2\phi,\\
%\theta_{\rm c}&\rightarrow \theta_{\rm c}-\phi,
%\end{align*}
in Eq.~\eqref{input_d}, resulting in $\boldsymbol{\sigma}_{{\rm in},\phi}$ and $\boldsymbol{d}_{{\rm in},\phi}$.

%%%%%%%%%%%%%%%%%%%%%%%%
%\subsection{Gaussian environment}
%%%%%%%%%%%%%%%%%%%%%%%%
%\textit{Gaussian environment ---.} 
We consider the Gaussian environment, under which the Gaussian probe state evolves, but still remains in a Gaussian state. The dynamics of the probe state $\hat{\rho}$ evolving under a typical Gaussian dissipative channel in thermal equilibrium can be described by the quantum master equation, written in the interaction picture as 
\begin{align}
\frac{d\hat{\rho}(t)}{dt}=
\frac{\gamma}{2} \left\{ n_{\rm e} {\cal L}[\hat{a}^{\dagger}] +(n_{\rm e}+1){\cal L}[ \hat{a}] \right\}\hat{\rho}(t),
%- \frac{\gamma}{2}\left\{m_{\rm e}^{*} {\cal D}[ \hat{a}] +m_{\rm e}{\cal D}[ \hat{a}^{\dagger}] \right\}\hat{\rho}(t),
\label{master}
\end{align}
where ${\cal L}[ \hat{o}]\hat{\rho}(t)=\left(2\hat{o}\hat{\rho}\hat{o}^{\dagger}-\hat{o}^{\dagger}\hat{o}\hat{\rho}-\hat{\rho}\hat{o}^{\dagger}\hat{o}\right)$ with a damping rate of $\gamma$, and $n_{\rm e}\in \mathbb{R}$ represents the average number of thermal photons of the environment~\cite{Ferraro05}. The terms proportional to ${\cal L}\left[\hat{a}\right]$ and to ${\cal L}\left[\hat{a}^{\dagger}\right]$ describe losses and phase-insensitive linear amplification processes, respectively.
%The positivity of the density matrix imposes the constraint $\vert m_{\rm e}\vert^{2}\le n_{\rm e}(n_{\rm e}+1)$. 
The solution of Eq.~\eqref{master} can be written for the covariance matrix and the first moment vector as $\boldsymbol{\sigma} = 
\left(1-\eta\right)\boldsymbol{\sigma}_{\infty}+\eta\boldsymbol{\sigma}_{\rm in}$, and $\boldsymbol{d} = \sqrt{\eta} \boldsymbol{d}_{\rm in}$,
%\begin{align*}
%\boldsymbol{\sigma} &= 
%\left(1-\eta\right)\boldsymbol{\sigma}_{\infty}+\eta\boldsymbol{\sigma}_{\rm in},\\
%\boldsymbol{d} &= \sqrt{\eta} \boldsymbol{d}_{\rm in},
%\end{align*}
where $\eta=e^{-\gamma t}$ denotes the effective transmission coefficient and $\boldsymbol{\sigma}_{\infty}=(n_{\rm e}+\frac{1}{2})\mathbb{I}_{2}$ with $\mathbb{I}_{2}$ being a $2\times2$ identity matrix. 
%\begin{align*}
%\boldsymbol{\sigma}_{\infty}
%=
%%\begin{pmatrix}
%%(n_{\rm e}+\frac{1}{2})& \Im[m_{\rm e}]\\
%%\Im[m_{\rm e}] &(n_{\rm e}+\frac{1}{2})-\Re[m_{\rm e}]
%%\end{pmatrix}.
%\begin{pmatrix}
%(n_{\rm e}+\frac{1}{2})& 0\\
%0 &(n_{\rm e}+\frac{1}{2})
%\end{pmatrix}.
%\end{align*}
Note that the output state characterized by $\boldsymbol{\sigma}$ and $\boldsymbol{d}$ is still a Gaussian state~\cite{Weedbrook12}. 
%In this work, we concentrate on an environment in thermal \textcolor{red}{equilibrium, and} $n_{\rm e}$ coincides with the average number of thermal photons in the environment. 
The evolution of the state $\hat{\rho}$ under such thermal environment commutes with the phase shift operation introduced above, so that all losses present in the channel can be assumed to have occurred before the phase shifter, causing a modification to parameters in Eq.~\eqref{input}. Consequently, the state that contains the effect of losses is written in the same decomposition of Eq.~\eqref{input}, but with modified parameters given as
\begin{widetext}
\begin{align}
\alpha_{\rm in}&\rightarrow ~~~\alpha=\sqrt{\eta}\alpha_{\rm in},\label{modifiedalpha}\\
r_{\rm in}&\rightarrow ~~~r=\frac{1}{2}{\rm ln}\left[\frac{(1-\eta)(1+2n_{\rm e})+\eta(1+2n_{\rm th,in})e^{2r_{\rm in}}}{\sqrt{\left[\eta(1+2n_{\rm th,in})+(1-\eta)(1+2n_{\rm e})\right]^{2}+4\eta(1-\eta)(1+2n_{\rm th,in})(1+2n_{\rm e})\sinh^{2}r_{\rm in}}}\right],\label{modifiedr}\\
n_{\rm th,in}&\rightarrow n_{\rm th}=\frac{1}{2}\sqrt{\left[\eta(1+2n_{\rm th,in})+(1-\eta)(1+2n_{\rm e})\right]^{2}+4\eta(1-\eta)(1+2n_{\rm th,in})(1+2n_{\rm e})\sinh^{2}r_{\rm in}}-\frac{1}{2},\label{modifiednth}
\end{align}
\end{widetext}
for
\begin{align}
\hat{\rho}&=\hat{D}(\alpha)\hat{S}(\xi)\hat{\rho}_{T}(n_{\rm th})\hat{S}^\dagger(\xi)\hat{D}^\dagger(\alpha),
\label{lossy_input}
\end{align}
where $\xi=re^{i\theta_{\rm s}}$ and the modified thermal state $\hat{\rho}_{T}(n_{\rm th})$ has the average photon number of $n_{\rm th}$. Although the modified parameters in Eq.~\eqref{lossy_input} represent all kinds of single-mode Gaussian states, it is worth using the expressions in Eqs.~\eqref{modifiedalpha} to  \eqref{modifiednth} to distinguish the role of the initial thermal photons ($n_{\rm th,in}$) from that of the environmental thermal photons ($n_{\rm e}$). Also note that the initial phases $\theta_{\rm c}$ and $\theta_{\rm s}$ remain the same due to the fact that a thermalization process does not affect the phase of the system. As mentioned, the phase shift operation is considered to occur to this lossy state, and in short, the probe state is transformed as
\begin{align}
\boldsymbol{\sigma}_{\rm in}\xrightarrow[]{\text{loss}}\boldsymbol{\sigma}\xrightarrow[]{\text{phase shift}}\boldsymbol{\sigma}_{\phi},\\
\boldsymbol{d}_{\rm in}\xrightarrow[]{\text{loss}}\boldsymbol{d}\xrightarrow[]{\text{phase shift}}\boldsymbol{d}_{\phi}.
\end{align}
We then analyze the output state of $\boldsymbol{\sigma}_{\phi}$ and $\boldsymbol{d}_{\phi}$ by a Gaussian measurement, which we shall introduce below.

%%%%%%%%%%%%%%%%%%%%%%%%
%\subsection{Gaussian measurement}
%%%%%%%%%%%%%%%%%%%%%%%%
%\textit{Gaussian measurement ---.}
The POVM element yielding a measurement outcome $y$ from a general Gaussian measurement can be written as
\begin{align}
\hat{\Pi}_y=\frac{1}{\pi}\hat{D}(y)\hat{\Pi}^{0}\hat{D}^\dagger(y),
\label{Gaussian_measurement}
\end{align}
where $\hat{\Pi}^{0}$ is a density matrix of a single-mode Gaussian state~\cite{Weedbrook12, giedke2002}. 
The probability of obtaining the measurement outcome $y$ is calculated by the overlap between the phase-encoded probe state $\hat{\rho}_{\phi}$ and the displaced measurement basis $\hat{\Pi}_{y}$, i.e., $p(y)=\text{Tr}[\hat{\Pi}_{y}\hat{\rho}_{\phi}]$. The displacement operator $\hat{D}(y)$ varies the center of the measurement basis to scan across the entire phase space, so that $\int dy \hat{\Pi}_{y}=\mathbb{I}$.
Note that the probability distribution of the measurement outcome for $\hat{\Pi}^{0}$ being a squeezed thermal state can be decomposed into a mixture of those for $\hat{\Pi}^{0}$ being squeezed vacuum states. We thus assume $\hat{\Pi}^{0}$ to be only the squeezed vacuum state without loss of generality according to the data processing inequality~\cite{Zamir98,Ferrie14}. Typical types of Gaussian measurement are the homodyne measurement [shown in Fig.~2(a)] and heterodyne measurement [shown in Fig.~2(b)], for which $\hat{\Pi}^{0}$ is an infinitely squeezed vacuum state, and $\hat{\Pi}^{0}$ is a vacuum state, respectively. Such general Gaussian POVMs can be performed experimentally by using general-dyne measurement~\cite{genoni2014}, as shown in Fig.~2(b). The squeezing parameter of $s e^{i\psi}$ with $s\ge0$, characterizing $\hat{\Pi}^{0}$, can be controlled in the general-dyne measurement setup by adjusting a transmittance $\tau$ of the beam splitter in a setup shown in Fig.~2(b), i.e., $s=\ln\sqrt{\tau/(1-\tau)}$ with $\tau\ge1/2$, and the phase $\psi$ can be tuned by varying phases of the local oscillator modes in sub-homodyne detection setups. The outcome $y$ is then obtained as~\cite{genoni2014}
\begin{align}
y
%&=\frac{X_{\psi/2}}{\sqrt{2\tau}}\cos\frac{\psi}{2}-\frac{P_{\psi/2}}{\sqrt{2(1-\tau)}}\sin\frac{\psi}{2} \nonumber \\ 
%&\quad+i~\bigg(\frac{X_{\psi/2}}{\sqrt{2\tau}}\sin\frac{\psi}{2}+\frac{P_{\psi/2}}{\sqrt{2(1-\tau)}}\cos\frac{\psi}{2}\bigg)\nonumber \\
&=\frac{1}{\sqrt{2\tau}} X_{\psi/2}e^{i\frac{\psi}{2}}+\frac{i}{\sqrt{2(1-\tau)}}P_{\psi/2}e^{i\frac{\psi}{2}},
\end{align}
where $X_{\psi/2}$ and $P_{\psi/2}$ are the rotated quadrature variables, being measured in the respective output ports of the beam splitter. 

\begin{figure}[b]
\centering
\includegraphics[width=0.46\textwidth]{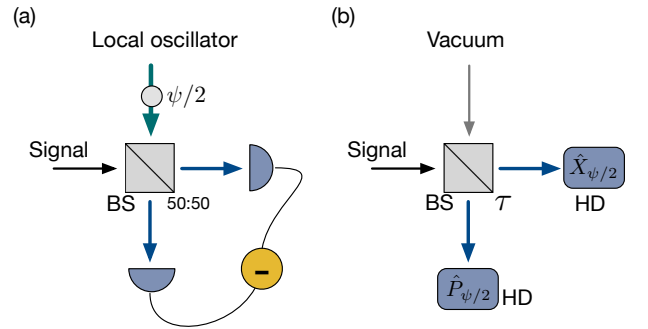}
\caption{Optical setups of Gaussian measurements.
(a) Setup for homodyne detection, where the signal and the local oscillator with phase $\psi/2$ are mixed by a 50:50 beam splitter and the difference of photocurrents is measured. This setup implements the measurement of the rotated quadrature operator $\hat{X}_{\psi/2}$.
(b) Measurement setup for general Gaussian measurement characterized by a squeezing parameter $se^{i\psi}$. The signal and the vacuum are mixed by a beam splitter with a transmittance $\tau$, and quadrature operators $\hat{X}_{\psi/2}$ and $\hat{P}_{\psi/2}$ are measured by the homodyne detection on the respective output modes.}
\label{m_setup}
\end{figure}

%%%%%%%%%%%%%%%%%%%%%%%%%
%%%%%%%%%%%%%%%%%%%%%%%%%
%\section{Estimation error}
%%%%%%%%%%%%%%%%%%%%%%%%%
%%%%%%%%%%%%%%%%%%%%%%%%%
%Here we introduce the figure of merit, the lower bound of an estimation error, which we use for assessing our phase estimation. 

%%%%%%%%%%%%%%%%%%%%%%%%
%\subsection{Cram\'er-Rao bound}
%%%%%%%%%%%%%%%%%%%%%%%%
%\textit{Cram\'er-Rao bound ---.}
From a parameter estimation theory, the error of the estimator $\hat{\phi}$ is typically defined by the mean-squared-error $\Delta^2 \phi=\langle(\hat{\phi}-\phi)^2\rangle$, where $\langle .. \rangle$ denotes the average taken over all measurement results and $\phi$ is the true value of the parameter. It is known that for any unbiased estimator, the error $\Delta^2 \phi$ is bounded by the inverse of FI, written by
\begin{equation}
\Delta^2 \phi \ge \frac{1}{MF(\phi)},
\label{CRineq}
\end{equation}
where $M$ denotes the number of repetition of measurement and the FI is defined as
\begin{align}
F(\phi)=\int dy \frac{1}{p(y\vert \phi)}\bigg(\frac{\partial p(y\vert \phi)}{\partial \phi}\bigg)^2.
\label{FI}
\end{align}
Here, $p(y\vert \phi)dy$ is a conditional probability of finding the experimental result between $y$ and $y+dy$ for a given parameter $\phi$. Inequality~\eqref{CRineq} is called the Cram\'{e}r-Rao inequality~\cite{cramer1946}, and can be asymptotically saturated in the limit of large $M$ by the maximum likelihood estimator~\cite{Fisher25}.

The Gaussian measurement of Eq.~\eqref{Gaussian_measurement} we consider in this work, by definition, produces a Gaussian probability distribution for the measurement outcomes $y$'s. In this case, the FI can be calculated in terms of the second moment matrix $\Sigma$ and the first moment vector $\nu$ of the measurement outcome probability distribution via~\cite{Kay93,Porat86}
\begin{align}
F(\phi)=\frac{\partial \nu^{\text{T}}}{\partial \phi}\Sigma^{-1}\frac{\partial \nu}{\partial \phi}+\frac{1}{2}\text{Tr}\left[\Sigma^{-1}\frac{\partial\Sigma}{\partial\phi}\Sigma^{-1}\frac{\partial\Sigma}{\partial\phi}\right].
\label{FI_gaussian}
\end{align}
In the case of a general Gaussian measurement, there are free parameters that need to be optimized to maximize $F(\phi)$: the squeezing parameters of $s$ and $\psi$ for $\hat{\Pi}_0$.

%%%%%%%%%%%%%%%%%%%%%%%%
%\subsection{Quantum Cram\'er-Rao bound}
%%%%%%%%%%%%%%%%%%%%%%%%
%\textit{Quantum Cram\'er-Rao bound ---.}
The Cram\'er-Rao bound provides the ultimate bound for a chosen measurement setup, but there is no guarantee that the chosen measurement setting is optimal. In other words, the FI of Eq.~\eqref{FI} varies with measurements and is maximized by choosing an optimal measurement. The optimization is done over all POVMs such that $\hat{\Pi}_k\geq 0$ and $\int dk \hat{\Pi}_k=\mathbb{I}$, yielding the maximal FI as
\begin{align}
H(\phi)=\max_{\{\hat{\Pi}_k\}}F(\phi),
\label{QFI}
\end{align}
called the QFI~\cite{Braunstein94,Braunstein96}. Thus, the QFI gives the ultimate lower bound of the mean-squared-error, written as
\begin{align}
\Delta^2\phi\ge\frac{1}{MF(\phi)}\ge\frac{1}{MH(\phi)}.
\label{QCRB}
\end{align}
This last expression is called the quantum Cram\'{e}r-Rao inequality.

For a given density matrix $\hat{\rho}=\sum_n p_n \ket{\psi_n}\bra{\psi_n}$, where $\langle\psi_n\vert\psi_m\rangle=\delta_{n,m}$, evolving to $\hat{\rho}_\phi=e^{-i\hat{G}\phi}\hat{\rho}e^{i\hat{G}\phi}$ with a generator $\hat{G}$, the QFI can be calculated as~\cite{Paris09}
\begin{align}
H(\phi)=2\sum_{n,m}\frac{(p_n-p_m)^2}{p_n+p_m}|\langle\psi_n|\hat{G}|\psi_m\rangle|^2.
\end{align}
In our case, the generator is given as $\hat{G}=\hat{a}^\dagger\hat{a}$, and the QFI is thus found to be~\cite{Pinel13,Jiang14}
\begin{align} 
H(\phi)=&\frac{2(2n_\text{th}+1)^2 \sinh^2{2r}}{2n_\text{th}^2+2n_\text{th}+1}\nonumber\\
&\quad+\frac{4\vert\alpha\vert^2}{2n_\text{th}+1}\left\vert\cosh r-e^{i(\theta_{\rm s}-2\theta_{\rm c})}\sinh r\right\vert^2.
\label{QFI_for_gaussian_state}
\end{align}
Note that there is no dependence of $\phi$, so that the ultimate error bound is equal for all $\phi$'s.

%%%%%%%%%%%%%%%%%%%%%%%%
%%%%%%%%%%%%%%%%%%%%%%%%
%\section{Optimal Gaussian measurements}\label{Section_Result}
%\section{Optimal Gaussian measurements}\label{Section_Result}
\section{Results}\label{Section_Result}
%%%%%%%%%%%%%%%%%%%%%%%%
%%%%%%%%%%%%%%%%%%%%%%%%
\subsection{Optimal Gaussian measurements}
We look for an optimal Gaussian measurement setup for the phase estimation with single-mode Gaussian probe states. A Gaussian measurement is said to be an optimal Gaussian measurement if it is optimized to yield a maximal FI. Furthermore, we call it the optimal measurement if the maximized FI reaches the QFI obtainable by an optimal POVM. For single-mode Gaussian probe states classified to three types, we explore whether the optimal Gaussian measurement schemes can constitute the optimal measurement setup.

%%%%%%%%%%%%%%%%%%%%%%%%
%\subsection{Displaced thermal state}\label{Section_DTS}
%%%%%%%%%%%%%%%%%%%%%%%%
\textit{Displaced thermal state ---.}\label{Section_DTS}
Let us first consider a displaced thermal state (DTS) of Eq.~\eqref{input} with $r_{\rm in}=0$ in a lossy channel characterized by $\eta$ and $n_{\rm e}$. The modified parameters of Eqs.~\eqref{modifiedalpha}-\eqref{modifiednth} due to loss are given by
\begin{align}
r&=0,\\
n_{\rm th}&=\frac{1}{2}\left[\eta(2n_{\rm th,in}+1)+(1-\eta)(2n_{\rm e}+1)-1\right],\label{nDTS}\\
\alpha&=\sqrt{\eta}\alpha_{\rm in}\label{alphaDTS}.
\end{align}
For the DTS, the QFI of Eq.~\eqref{QFI_for_gaussian_state} takes the form of~\cite{Aspachs09}
\begin{align}
H_{\rm DTS}=\frac{4\abs{\alpha}^{2}}{2n_{\rm th}+1},
\label{HDTS}
\end{align}
whereas the FI for a general Gaussian measurement is written as
\begin{align}
F_{\rm DTS}=\frac{2\abs{\alpha}^{2}\left[1+2n_{\rm th}+\cosh2s_{\rm DTS}-\cos\chi_{\rm DTS}\sinh 2s_{\rm DTS}\right]}{1+2n_{\rm th}(n_{\rm th}+1)+(2n_{\rm th}+1)\cosh2s_{\rm DTS}},
\label{FDTS}
\end{align}
where $\chi_{\rm DTS}=2(\theta_{\rm c}-\phi)-\psi$. We find that when $s_{\rm DTS}\rightarrow\infty$ and $\chi_{\rm DTS}=\pi$, $F_{\rm DTS}$ is the same as $H_{\rm DTS}$, and when $n_{\rm th}=0$, $F_{\rm DVS}=H_{\rm DVS}=4\abs{\alpha}^{2}$. This means that the homodyne detection is the optimal measurement setup for any $n_{\rm th}$, $\alpha$, $\eta$, and $n_{\rm e}$. More detailed behaviors are explained below. 
\begin{figure}
\centering
\includegraphics[width=0.48\textwidth]{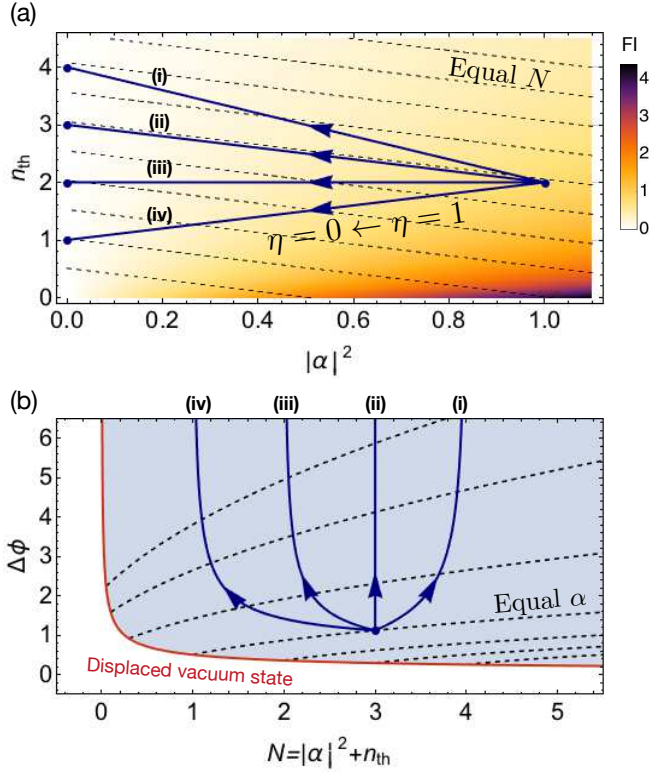}
\caption{Phase estimation with DTSs and optimal Gaussian measurements.
(a) The density plot represents $F_{\rm DTS}$ of Eq.~\eqref{FDTS} as a function of $\abs{\alpha}^{2}$ and $n_{\rm th}$. The dashed lines show the cases of the states with an equal average photon number. Four examples of Gaussian environments are considered here for a given input state of $\abs{\alpha_{\rm in}}^{2}=1$ and $n_{\rm th,in}=1$:  (i) $n_{\rm e}=N_{\rm in}+2$, (ii) $n_{\rm e}=N_{\rm in}$, (iii) $n_{\rm e}=n_{\rm th,in}$, and (iv) $n_{\rm e}=n_{\rm th,in}-1$. The direction of the arrows is along with a decrease in the transmission coefficient $\eta$. (b) The Cram\'er-Rao bound $\Delta \phi$ is shown in terms of the averaged photon number $N$ of the state arriving at the measurement setup for any $\vert\alpha\vert$ and $n_{\rm th}$. The shaded region represents all possible errors for any combination of $\vert\alpha\vert$ and $n_{\rm th}$ that builds up the photon number $N$ considered. The lower bound of the shaded region is given by the case using the displaced vacuum state, and the dashed lines show the cases of the states with an equal $\vert\alpha\vert$. The four examples of (i)-(iv) considered in (a) are also considered in (b), showing the error bounds for all cases grow up so rapidly.}
\label{DTS}
\end{figure}

In Fig.~3(a), the density plot represents $F_{\rm DTS}$ as a function of $\abs{\alpha}^{2}$ and $n_{\rm th}$, manifesting that the largest FI is obtained at the right lower corner, where $n_{\rm th}=0$, given the parameter regime. The dashed lines correspond to the states with an equal average photon number $N = \abs{\alpha}^{2}+n_{\rm th}$. This shows that for a given average photon number, reducing thermal contributions enables to achieve larger FIs. The effects of loss channels are also considered here for a given example input state of $\abs{\alpha_{\rm in}}^{2}=1$ and $n_{\rm th,in}=1$ (i.e., $N_{\rm in}=2$) when (i) $n_{\rm e}>N_{\rm in}$, (ii) $n_{\rm e}=N_{\rm in}$, (iii) $n_{\rm e}=n_{\rm th,in}$, and (iv) $n_{\rm e}<n_{\rm th,in}$. The arrows represent a decrease in the transmittance coefficient $\eta$ (or equivalently an increase in the loss rate $\gamma$ for a given propagation time $t$), changing $n_{\rm th}$ and $\alpha$ according to Eqs.~\eqref{nDTS} and \eqref{alphaDTS}. It clearly reveals that all lossy cases decrease the FI with $\eta$, i.e., losses are detrimental. 

The estimation error bound $\Delta \phi$ is also shown as a function of an average photon number $N$ in Fig.~3(b), for any possible $\alpha$ and $n_{\rm th}$. It displays that the minimal error is achieved only by the displaced vacuum state, i.e., a pure coherent state $\ket{\alpha}$. The dashed lines represent the states with an equal $\alpha$ but $n_{\rm th}$ varying, i.e., indicating that adding thermal photons to the probe state always increases the estimation error. It is also shown that the corresponding errors to examples of (i)-(iv) considered in Fig.~3(a) shoot up so quickly. 

Therefore, the best state out of all possible displaced thermal states for a fixed average photon number is a pure coherent state. This conclusion continues to hold even when losses are present, i.e., the use of a pure coherent state with $n_{\rm th,in}=0$ as an input attains the ultimate limit obtained by the QFI for any $n_{\rm e}$ and $\eta$. For any cases, this ultimate limit is achieved by the homodyne detection, one of typical Gaussian measurements. In other words, the optimal Gaussian measurement is the optimal measurement for the case when the displaced thermal state is used for phase estimation. 

%%%%%%%%%%%%%%%%%%%%%%%%
%\subsection{Squeezed thermal state}\label{Section_STS}
%%%%%%%%%%%%%%%%%%%%%%%%
\textit{Squeezed thermal state ---.}\label{Section_STS}
A second type of single-mode Gaussian states is a squeezed thermal state (STS) in Eq.~\eqref{input} with $\alpha_{\rm in}=0$. 
Consideration of such state is important when impure squeezed states are used in an experiment~\cite{Berni2015}. Even much highly squeezed states that have recently been generated~\cite{Vahlbruch2016} have a non-negligible thermal noise, causing an asymmetry between the squeezing and anti-squeezing level in units of dB.
In the presence of loss, the QFI of Eq.~\eqref{QFI_for_gaussian_state} takes the form of~\cite{Aspachs09}
\begin{align}
H_{\rm STS}=C_{H}\sinh^{2}2r,
\label{HSTS}
\end{align}
where $C_{H}=2(2n_{\rm th}+1)^{2}/(2n_{\rm th}^{2}+2n_{\rm th}+1)$, and the modified parameters of $r$ and $n_{\rm th}$ are to be obtained by Eqs.~\eqref{modifiedalpha} and~\eqref{modifiednth}. Here, $H_{\rm STS}$ reveals a remarkable positive contribution of thermal photons of the probe state~\cite{Aspachs09, Safranek16}; a twofold enhancement in the QFI is asymptotically achieved when $n_{\rm th}\rightarrow\infty$~\cite{Safranek16}. 
It is also interesting to see that for a given total energy $N$, $H_{\rm STS}$ is greater than the standard quantum limit (SQL) of $H_{\rm SQL}=4N$ when $\sinh^{2}r > \{ 2n_{\rm th}^{2}- 2n_{\rm th}-1+[1+4n_{\rm th}(n_{\rm th}+1)(n_{\rm th}^{2}+n_{\rm th}+3)]^{1/2}\} /4(2n_{\rm th}+1)$. This condition for quantum enhancement can be shown to be stricter than the non-classicality condition of STSs, written as $e^{-2r}(2n_{\rm th}+1)>1$~\cite{Marian1993,Asboth2005}, since in phase estimation a pure coherent state is definitely superior to a mixture of coherent states, into which STSs can be decomposed when the non-classicality condition is violated.
%It is because when the non-classicality condition is violated, the STS can be considered as a mixture of coherent states with different amplitudes, and so definitely inferior to a pure coherent state in phase estimation. 

\begin{figure}[b]
\centering
\includegraphics[width=0.46\textwidth]{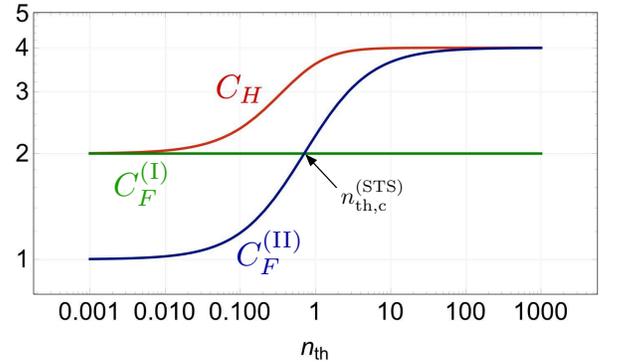}
\caption{
Comparison of the prefactor $C$'s. The prefactor $C_{\rm H}$ for the QFI is by definition always the largest, but the $C_{\rm F}^{\rm (I)}$ and $C_{\rm F}^{\rm (II)}$ are rather competitive: The first type ($C_{\rm F}^{\rm (I)}$) is the optimal Gaussian measurement when $n_{\rm th}\le n_{\rm th,c}^{\rm (STS)}\equiv 2^{-1/2}$, while the second type ($C_{\rm F}^{\rm (II)}$) is the optimal Gaussian measurement when $n_{\rm th}\ge n_{\rm th,c}^{\rm (STS)}$. Note that none of them is the same as the QFI, but $C_{\rm F}^{\rm (I)}$ ($C_{\rm F}^{\rm (II)}$) can asymptotically be similar to $C_{\rm H}$ in the limit of small (large) $n_{\rm th}$. In other words, the optimal Gaussian measurements are nearly optimal setups in those limits.}
\label{comparison}
\end{figure}

For a squeezed vacuum state (SVS) in the absence of loss (i.e., $r=r_{\rm in}$ and $n_{\rm th}=0$), the homodyne detection is known to be an optimal measurement~\cite{monras2006, olivares2009}, i.e., the FI of Eq.~\eqref{FI_gaussian} in the limit $s_{\rm SVS}\rightarrow\infty$ is written as
\begin{align}
F_{\rm SVS}=2\sinh^{2} 2r,
\label{FSVS}
\end{align}
where the optimal angle is chosen such that $\cos\chi_{\rm SVS}=\tanh 2r$ for $\psi=\theta_{\rm s}-2\phi-\chi_{\rm SVS}$~\cite{Aspachs09,olivares2009,Berni2015}. It is apparent that the FI of Eq.~\eqref{FSVS} is the same as the QFI when $n_{\rm th}=0$, i.e., the optimal Gaussian measurement is the optimal measurement when a squeezed vacuum probe state is used in the absence of losses. 

When a thermal noise is initially present in the input state or flows into the probe state from the environment, i.e., $n_{\rm th}\neq0$, a general Gaussian measurement needs to be optimized to maximize the FI. As a result, we find two types of optimal Gaussian measurements depending on the value of $n_{\rm th}$. The first type is achieved in the limit $s_{\rm STS}^{\rm (I)}\rightarrow\infty$ with $\cos\chi_{\rm STS}^{\rm (I)}=\tanh 2r$, while the second type is when $s_{\rm STS}^{\rm (II)}=r$ with $\cos\chi_{\rm STS}^{\rm (II)}=1$. The corresponding FIs are written as
\begin{align}
F_{\rm STS}^{\rm (I)}&=C_{F}^{\rm (I)}\sinh^{2} 2r,\label{FSTSinfinity}\\
F_{\rm STS}^{\rm (II)}&=C_{F}^{\rm (II)}\sinh^{2}2r,\label{FSTSr}
\end{align}
where $C_{F}^{\rm (I)}=2$, and $C_{F}^{\rm (II)}=\left[(2n_{\rm th}+1)/(n_{\rm th}+1)\right]^{2}$, respectively. We then compare the prefactor $C$'s in terms of the thermal photon number. Figure~4 shows that when $n_{\rm th}<n_{\rm th,c}^{\rm (STS)}\equiv 2^{-1/2}$, $C_{F}^{\rm (I)}$ outperforms $C_{F}^{\rm (II)}$, but the relative behavior is reversed when $n_{\rm th}>n_{\rm th,c}^{\rm (STS)}$. At $n_{\rm th}=n_{\rm th,c}^{\rm (STS)}$, they are the same, i.e., $C_{F}^{\rm (I)}=C_{F}^{\rm (II)}$. This means that the homodyne detection is the optimal Gaussian measurement when $n_{\rm th}\le n_{\rm th,c}^{\rm (STS)}$, while the second type Gaussian measurement is the optimal Gaussian measurement when $n_{\rm th}\ge n_{\rm th,c}^{\rm (STS)}$.
This is in stark contrast to the conclusion of the work in Ref.~\cite{Monras2013}, where a homodyne detection is found to be always optimal among Gaussian measurements for the case that first moments are fixed. Such notable disagreement occurs since the proof given in Ref.~\cite{Monras2013} has not taken into account truly all Gaussian measurements, but only the Gaussian measurements that project the input state into mixed Gaussian states. The latter misses the optimality of the above type-II Gaussian measurement that outperforms the homodyne detection when $n_{\rm th}>n_{\rm th,c}^{\rm (STS)}$.
In the limit of small or large $n_{\rm th}$, the FI with an optimally chosen Gaussian measurement is asymptotically close to the QFI, but not equal. Therefore, the Gaussian measurement settings provide nearly optimal measurement setups in the limit of small or large $n_{\rm th}$.

\begin{figure}
\centering
\includegraphics[width=0.48\textwidth]{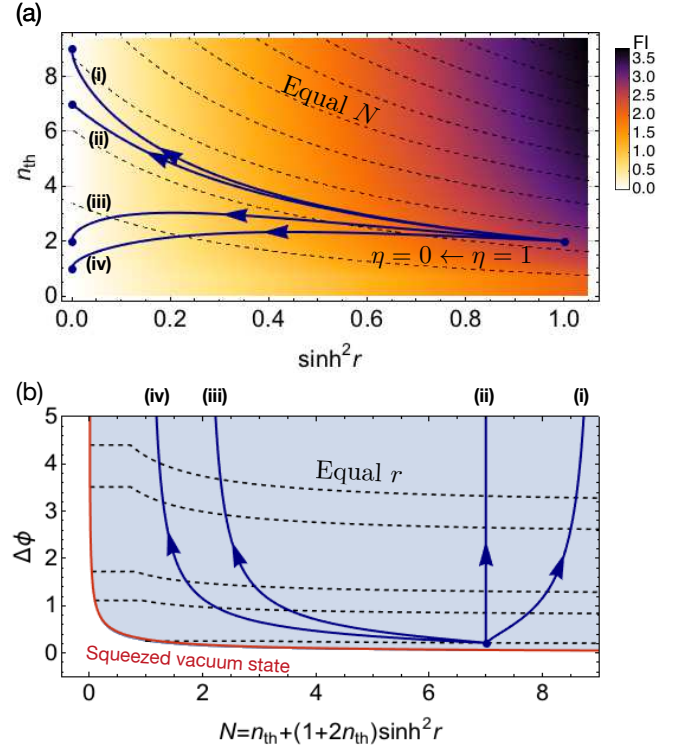}
\caption{
Phase estimation with STSs and optimal Gaussian measurements.
(a) $F_{\rm STS}$ is shown as a function of $\sinh^{2}r$ and $n_{\rm th}$. In the region, where $n_{\rm th}\le2^{-1/2}$, the first type Gaussian measurement is employed, while the second type Gaussian measurement is used for the other region. The dashed lines denote the states having an equal average photon number. Four examples of Gaussian environments are considered here for a given input state of $\sinh^{2}r_{\rm in}=1$ and $n_{\rm th,in}=2$:  (i) $n_{\rm e}=N_{\rm in}+7$, (ii) $n_{\rm e}=N_{\rm in}$, (iii) $n_{\rm e}=n_{\rm th,in}$, and (iv) $n_{\rm e}=n_{\rm th,in}-1$. The transmission coefficient $\eta$ decreases in the direction of the arrow. (b) The Cram\'er-Rao inequality provides the lower estimation error bound $\Delta \phi$, shown in terms of the average photon number $N$ of the state arriving at the measurement setup for any $r$ and $n_{\rm th}$. The shaded region covers all possible error bounds for any combination of $r$ and $n_{\rm th}$ that builds up the photon number $N$ considered. The shaded region is lower bounded by the case using the squeezed vacuum state, and the dashed lines show an equal $r$. The four examples of (i)-(iv) considered in (a) are also presented in (b), showing a rapid growth of the error bounds for all cases.}
\label{STS}
\end{figure}

In Fig.~5, we present detailed behaviors of phase estimation using STSs. In Fig.~5(a), the density plot represents the optimized FI in terms of $\sinh^{2}r$ and $n_{\rm th}$, showing that the largest FI is achieved at the upper right corner, in which both $\sinh^{2}r$ and $n_{\rm th}$ are maximal given the parameter regime. In other words, adding initial thermal photons ($n_{\rm th,in}$) with fixing a squeezing parameter helps to increase the FI, as in the QFI. Similar positive contributions of thermal photons have been reported in Refs.~\cite{Aspachs09, Safranek16}. However, when the total average photon number is fixed, which is often restricted when a vulnerable bio-chemical transducer is employed~\cite{Taylor16}, a pure squeezed state is required for a maximal FI. This is manifested by the dashed lines that denote the squeezed thermal states having an equal average photon number $N$. We also consider the effect of loss channel for a given example input state of $\sinh^{2}r_{\rm in}=1$ and $n_{\rm th,in}=2$ (i.e., $N_{\rm in}=7$) when (i) $n_{\rm e}>N_{\rm in}$, (ii) $n_{\rm e}=N_{\rm in}$, (iii) $n_{\rm e}=n_{\rm th}$, and (iv) $n_{\rm e}<n_{\rm th}$. As before, the arrows represent the direction along which the transmission coefficient $\eta$ decreases, or equivalently the loss rate $\gamma$ increases. It is clear that the FI monotonically decreases with a decrease of $\eta$ for any cases. Also note that unlike the initial thermal photons, the contribution of the environmental thermal photons ($n_{\rm e}$) is always negative.

In Fig.~5(b), the estimation error bound of the phase estimation using STSs is presented, in which the shaded region includes all possible values of error bounds for the considered states. The region is bounded by the lower limit, achieved by the case using the SVS. The dashed lines represent the states having an equal squeezing strength $r$. This shows that an increase of $n_{\rm th}$ while keeping $r$ unaltered helps to further decrease the estimation error, as already remarked previously. The lossy cases considered in Fig.~5(a) are also presented, displaying that the errors quickly shoot up with $\eta$.

%%%%%%%%%%%%%%%%%%%%%%%%
%\subsection{Displaced squeezed thermal state}\label{Section_DSTS}
%%%%%%%%%%%%%%%%%%%%%%%%
\textit{Displaced squeezed thermal state ---.}\label{Section_DSTS}
We finally consider the most general single-mode Gaussian state given in Eq.~\eqref{input}, i.e., a displaced squeezed thermal state (DSTS) that contains displacement, squeezing, and thermal photons. For such a general state, the QFI of Eq.~\eqref{QFI_for_gaussian_state} is maximized with the optimal phase relation $\theta_{\rm c}=\theta_{\rm s}/2$. It has been shown that for a fixed average photon number $N$, $H_{\rm DSTS}$ is maximized when $\alpha=0$ and $n_{\rm th}=0$, i.e., the squeezed vacuum state is the optimal state yielding the maximal QFI~\cite{Pinel13}. One can also see that 
the behavior of QFI with increasing $n_{\rm th}$ turns around across points at which $2\alpha^{2}e^{2r}\sinh^{-2}2r = (2n_{\rm th}+1)^{3}/(1+2n_{\rm th}(n_{\rm th}+1))^{2}$, as in Ref.~\cite{Safranek16}.

Now we optimize Gaussian measurements in order to maximize the FI of Eq.~\eqref{FI_gaussian} for displaced squeezed thermal probe states. First of all, we set the optimal phase relations as $\psi=\theta_{\rm s}-2\phi-\chi$, and $\theta_{\rm c}=(\pi+\theta_{\rm s})/2$, which also covers the phase relations used above. Then, the optimal angle for $\chi$ needs to be found together with $s$ (measurement squeezing) for given $\alpha$, $r$, and $n_{\rm th}$.

Let us first consider the case, where no thermal photons are involved, i.e., $n_{\rm th}=0$, a displaced squeezed vacuum state (DSVS). Previously we have seen that the homodyne detection scheme provides the optimal measurement setup for both a displaced vacuum state and a squeezed vacuum state. One might then conclude that the homodyne detection would be the optimal measurement setup also for the displaced squeezed vacuum state. However, it is not the case as we discuss now. The FIs for the optimized Gaussian measurements we found are written as
\begin{align}
F_{\rm DSVS}^{\rm (I)}&=4e^{2r}\abs{\alpha}^{2},\label{FDSVS1}\\
F_{\rm DSVS}^{\rm (II)}&=\frac{\left[2\sinh2r+(1+\coth 2r)\abs{\alpha}^{2}\right]^{2}}{2},\quad\text{for}~r\neq0,\label{FDSVS2}
\end{align}
with $s_{\text{DSVS}}\rightarrow\infty$ (homodyne detection) for both cases, but different optimal angles of $\chi_{\rm DSVS}$ are chosen for given $\alpha$ and $r$ such that
\begin{align}
\cos\chi^{\rm (I)}_{\rm DSVS}&=1,\\
\cos\chi^{\rm (II)}_{\rm DSVS}&=\coth2r-\frac{2}{e^{2r} \abs{\alpha}^{2}+\sinh 4r},\label{OptimalPhaseDSVS2}
\end{align}
respectively. The above two types of optimal Gaussian measurements are complementary to each other: $F_{\rm DSVS}^{\rm (II)}$ is optimal when $\vert \cos\chi^{\rm (II)}\vert<1$ for $r\neq0$, while $F_{\rm DSVS}^{\rm (I)}$ is optimal when $\vert \cos\chi^{\rm (II)}\vert>1$. At the boundary, $F_{\rm DSVS}^{\rm (I)}=F_{\rm DSVS}^{\rm (II)}$. The condition of $\vert \cos\chi^{\rm (II)}\vert\le1$ can be reduced to $\abs{\alpha}\le \vert \tilde{\alpha}_{\rm max}^{\rm (DSVS)}\vert e^{-r}\sinh2r$ for $r\neq0$, where $\vert \tilde{\alpha}_{\rm max}^{\rm (DSVS)}\vert=\sqrt{2}$. Such homodyne detections are better than any other Gaussian measurements, but cannot be the optimal measurement that attains the QFI written as $H_{\rm DSVS}=2\sinh^{2}2r+4e^{2r}\abs{\alpha}^{2}$. One can also show that when $r=0$, $F_{\rm DSVS}^{\rm (I)}=4\abs{\alpha}^{2}$ is the same as $F_{\rm DVS}$, whereas when $\alpha=0$, $F_{\rm DSVS}^{\rm (II)}=2\sinh^{2}2r$ with Eq.~\eqref{OptimalPhaseDSVS2} being reduced to $\cos\chi^{\rm (II)}_{\text{DSVS}}=\tanh(2r)$ is equal to $F_{\rm SVS}$ of Eq.~\eqref{FSVS}.

Now let us turn to the case that thermal photons exist in the Gaussian probe state. For this general state, we find that three types of optimal Gaussian measurements exist and the corresponding FIs are written as
\begin{widetext}
\begin{align}
F_{\rm DSTS}^{\rm (I)}&=\frac{4 e^{2r}\abs{\alpha}^{2}}{2n_{\rm th}+1},\\
F_{\rm DSTS}^{\rm (II)}&=\frac{\left[2(2n_{\rm th}+1)\sinh2r+(1+\coth2r)\abs{\alpha}^{2}\right]^{2}}{2(2n_{\rm th}+1)^{2}},\quad\quad\quad\quad\quad\quad\quad\quad\quad\quad\quad\quad\quad\quad\quad\quad~\text{for}~r\neq 0,\\
F_{\rm DSTS}^{\rm (III)}&=\frac{(2n_{\rm th}+1)^{2}(2n_{\rm th}^{2}+2n_{\rm th}+1)\sinh^{2}2r +2n_{\rm th}(n_{\rm th}+1)(2n_{\rm th}+1)e^{2r}\abs{\alpha}^{2}}{2n_{\rm th}^{2}(n_{\rm th}+1)^{2}}\nonumber\\
&\quad\quad\quad\quad\quad\quad
-\frac{(2n_{\rm th}+1)^{3/2}\sinh2r \sqrt{(2n_{\rm th}+1)^{3} \sinh^{2}2r+4n_{\rm th}(n_{\rm th}+1) e^{2r}\abs{\alpha}^{2}}}{2n_{\rm th}^{2}(n_{\rm th}+1)^{2}},\quad~\text{for}~r\neq 0,
\end{align}
respectively. The respective optimal values of $s$ and $\chi$ are listed below.
\begin{itemize}
\item For type-I, $s_{\rm DSTS}^{\rm (I)}\rightarrow\infty~~\text{\&}~~\cos\chi_{\rm DSTS}^{\rm (I)}=0$,
\item For type-II, $s_{\rm DSTS}^{\rm (II)}\rightarrow\infty~~\text{\&}~~\cos\chi_{\rm DSTS}^{\rm (II)}=\frac{4(2n_{\rm th}+1)\sinh2r +2\coth2r(1+\coth2r) \abs{\alpha}^{2}}{ 4(2n_{\rm th}+1)\cosh2r+2(1+\coth2r)\abs{\alpha}^{2}\quad\quad\quad},\quad~\text{for}~r\neq 0$
\item For type-III, $s_{\rm DSTS}^{\rm (III)}=s_{\rm opt}~~\text{\&}~~\cos\chi_{\rm DSTS}^{\rm (III)}=0$,
\end{itemize}
where
\begin{align}
s_{\rm opt}&=
{\rm ln}\left[\left(
\frac{
(2n_{\rm th}+1) e^{4r}\abs{\alpha}^{2}
+(2n_{\rm th}+1)^{3/2}e^{2r}\sinh2r \sqrt{(2n_{\rm th}+1)^{3}\sinh^{2}2r+4n_{\rm th}(n_{\rm th}+1)e^{2r}\abs{\alpha}^{2}}
}
{(2n_{\rm th}+1)^{3}\sinh^{2}2r-e^{2r}\abs{\alpha}^{2}}
\right)^{1/2}\right] \quad~\text{for}~r\neq 0.
\end{align}
\end{widetext}

\begin{figure}[b]
\centering
\includegraphics[width=0.45\textwidth]{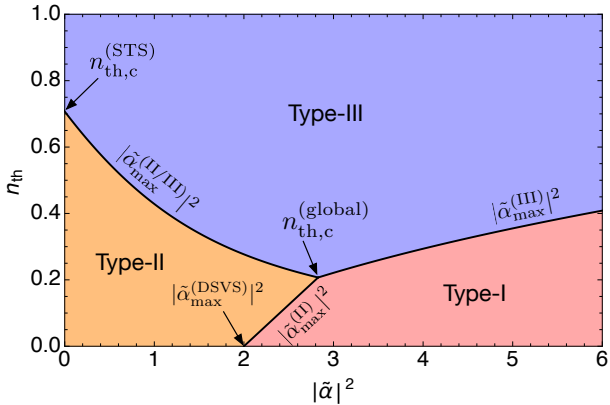}
\caption{
The regions, where one of optimal Gaussian measurements outperforms the others, are shown in terms of $n_{\rm th}$ and $\vert \tilde{\alpha}\vert^{2}=\vert\alpha\vert^{2}/e^{-2r}\sinh^{2}2r$. Solid lines represent the boundaries at which both intersecting types are optimal, and all of the three types are the optimal Gaussian measurement at a global critical point of $n_{\rm th,c}^{\rm (global)}$, where the boundaries coincide. When $n_{\rm th}=0$, a transition from type-II to type-I occurs at $\vert\tilde{\alpha}\vert=\vert \tilde{\alpha}_{\rm max}^{\rm (DSVS)}\vert$. When $\vert\tilde{\alpha}\vert=0$, a transition from type-II to type-III occurs at $n_{\rm th}=n_{\rm th,c}^{\rm (STS)}$. 
}
\label{Measurement_Phase}
\end{figure}

The condition $\vert \cos\chi_{\rm DSTS}^{\rm (II)}\vert\le1$ at which the type-II is available can be reduced to $\abs{\alpha}\le \vert \tilde{\alpha}_{\rm max}^{\rm (II)}\vert e^{-r}\sinh2r$ for $r\neq0$, where $\vert \tilde{\alpha}_{\rm max}^{\rm (II)}\vert=\sqrt{2(1+2n_{\rm th})}$ and this is the same as $\vert \tilde{\alpha}_{\rm max}^{\rm (DSVS)}\vert $ when $n_{\rm th}=0$. On the other hand, the condition for $s_{\rm opt}$ to be a positive real number is reduced to $\vert \alpha\vert<\vert\tilde{\alpha}_{\rm max}^{\rm (III)}\vert e^{-r}\sinh2r$ for $r\neq0$, where $\vert\tilde{\alpha}_{\rm max}^{\rm (III)}\vert=(1+2n_{\rm th})^{3/2}$. In addition, there exists another bound $\vert\tilde{\alpha}_{\rm max}^{\rm (II/III)}\vert$ to $\abs{\alpha}$, through which type-II and type-III are comparable. That is,  the FI for type-II is greater than that for type-III when $\vert\alpha\vert<\vert\tilde{\alpha}_{\rm max}^{\rm (II/III)}\vert e^{-r}\sinh2r$ for $r\neq0$, where $\vert\tilde{\alpha}_{\rm max}^{\rm (II/III)}\vert=\left((2n_{\rm th}+1)\left[1-(\sqrt{2}-1)(2n_{\rm th}+1)\right]\Big/n_{\rm th}\right)^{1/2}$, while the type-III outperforms the type-II when  $\vert\alpha\vert>\vert\tilde{\alpha}_{\rm max}^{\rm (II/III)}\vert e^{-r}\sinh2r$ for $r\neq0$. At the boundary, the FIs for type-II and type-III are the same. Interestingly, these three bounds coincide at $n_{\rm th}=n_{\rm th,c}^{\rm (global)}\equiv(\sqrt{2}-1)/\sqrt{2}$, i.e., resulting in $\vert \tilde{\alpha}_{\rm max}^{\rm (II)}\vert=\vert\tilde{\alpha}_{\rm max}^{\rm (III)}\vert=\vert\tilde{\alpha}_{\rm max}^{\rm (II/III)}\vert=2^{3/4}$, and the three types of measurements serve as an optimal Gaussian measurement. Although their setups are different, the FIs are the same at the global criticial point. Depending on the value of $\alpha$, $r$, and $n_{\rm th}$, there exist regions, where each of three types constitutes an optimal Gaussian measurement: 
\begin{itemize}
\item Type-I is an optimal Gaussian measurement when $\vert\tilde{\alpha}_{\rm max}^{\rm (III)}\vert\le\vert\tilde{\alpha}\vert$ for~$n_{\rm th}<n_{\rm th,c}^{\rm (global)}$, or when $\vert\tilde{\alpha}_{\rm max}^{\rm (II)}\vert\le\vert\tilde{\alpha}\vert$ for~$n_{\rm th}>n_{\rm th,c}^{\rm (global)}$. 
\item Type-II is an optimal Gaussian measurement when $\vert \tilde{\alpha}\vert\le\vert\tilde{\alpha}_{\rm max}^{\rm (II)}\vert$ for~$n_{\rm th}<n_{\rm th,c}^{\rm (global)}$, or when $\vert\tilde{\alpha}\vert\le\vert\tilde{\alpha}_{\rm max}^{\rm (II/III)}\vert$ for~$n_{\rm th}>n_{\rm th,c}^{\rm (global)}$.
\item Type-III is an optimal Gaussian measurement when $\vert\tilde\alpha_{\rm max}^{\rm (II)}\vert \le\vert\tilde\alpha\vert \le\vert\tilde\alpha_{\rm max}^{\rm (III)}\vert$ for~$n_{\rm th}>n_{\rm th,c}^{\rm (global)}$.
\item All of the three types are optimal Gaussian measurements at $n_{\rm th}=n_{\rm th,c}^{\rm (global)}$ and $\abs{\tilde{\alpha}}=2^{3/4}$.
\end{itemize}
These regions are clearly shown in Fig.~6. Particularly, in the limits $n_{\rm th}\ll1$, $n_{\rm th}\gg1$, $\abs{\tilde{\alpha}}\ll1$ or $\abs{\tilde{\alpha}}\gg1 $, the Gaussian measurement setups we found are nearly optimal setups, i.e., FI~$\approx$~QFI. Also note that this general distinction of the regions for the three types of optimal Gaussian measurements can be applied to all particular input states considered in the previous sections. For example, when $n_{\rm th}=0$, type-I Gaussian measurement leads to the FI of Eq.~\eqref{FDSVS1} and type-II Gaussian measurement results in the FI of Eq.~\eqref{FDSVS2}. When $\abs{\alpha}=0$, type-II Gaussian measurement gives rise to the FI of Eq.~\eqref{FSTSinfinity}, and type-III Gaussian measurement yields the FI of Eq.~\eqref{FSTSr}. When $r=0$, type-I Gaussian measurement leads to the FI of Eq.~\eqref{FDTS}. Such mapping from general three types to particular optimal Gaussian measurements is summarized in Table~\ref{mapping}.

\begin{table}
\centering
\includegraphics[width=0.48\textwidth]{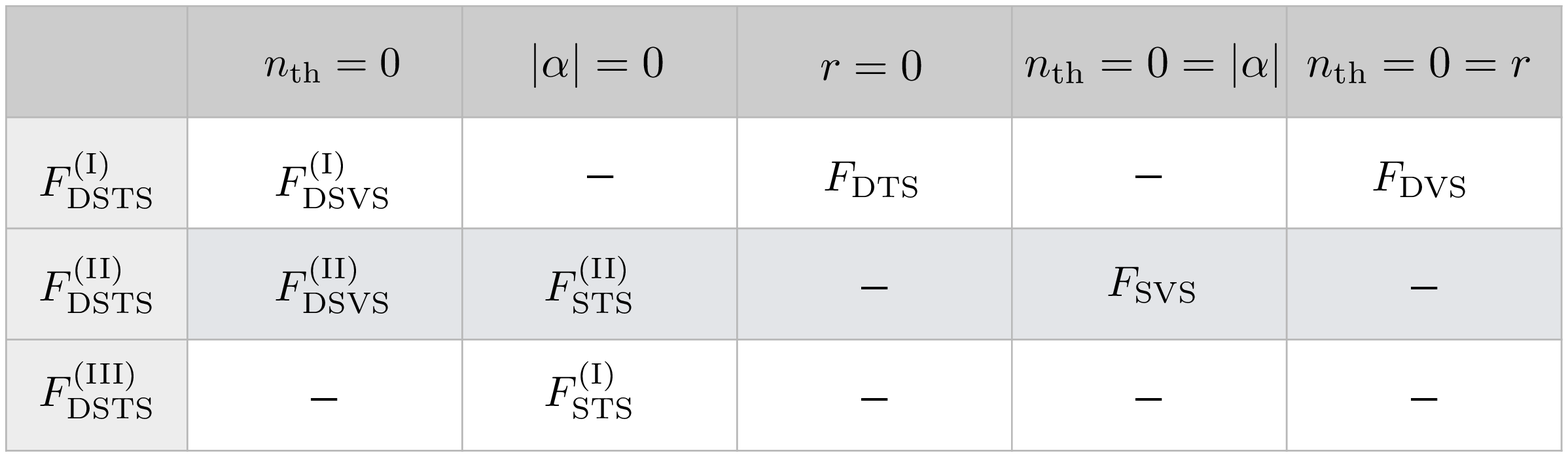}
\caption{
Reduction of three types of optimal Gaussian measurements to the measurements considered for particular Gaussian probe states is shown in terms of FI.}
%\setlength{\tabcolsep}{0.3em} % for the horizontal padding
%{\renewcommand{\arraystretch}{1}% for the vertical padding
%\begin{tabular}{ | c || c | c | c | c | c |}
%\hline
%Types & $n_{\rm th}=0$ & $\abs{\alpha}=0$ & $r=0$ & $n_{\rm th}=0=\abs{\alpha}$ & $n_{\rm th}=0=r$ \\ \hline\hline
%$F_{\rm DSTS}^{\rm (I)}$ & $F_{\rm DSVS}^{\rm (I)}$ & -- & $F_{\rm DTS}$ & -- & $F_{\rm DVS} $ \\ \hline 
%$F_{\rm DSTS}^{\rm (II)}$  & $F_{\rm DSVS}^{\rm (II)}$ & $F_{\rm STS}^{\rm (II)}$ & -- & $F_{\rm SVS}$ & -- \\ \hline
%$F_{\rm DSTS}^{\rm (III)}$  & -- & $F_{\rm STS}^{\rm (I)}$ & -- & -- & -- \\ \hline
%\end{tabular}
%}
%\caption{
%Reduction of three types of optimal Gaussian measurements to the measurements considered for particular Gaussian probe states is shown in terms of FI.}
\label{mapping}
\end{table}

%%%%%%%%%%%%%%%%%%%%%%%%
%%%%%%%%%%%%%%%%%%%%%%%%
\subsection{Optimal measurements beyond Gaussian measurements}
%%%%%%%%%%%%%%%%%%%%%%%%
%%%%%%%%%%%%%%%%%%%%%%%%
We have shown that for displaced thermal states and squeezed vacuum states, the optimized Gaussian measurements (i.e., the homodyne detection) constitute the optimal setup for the phase estimation, attaining the ultimate lower limit of estimation error. For the other kinds of single-mode Gaussian probe states, on the other hand, three types of optimized Gaussian measurements are found in general. However, the maximized FIs cannot exactly reach the QFI although they are nearly optimal in several limits. This means that a non-Gaussian measurement is required for those cases in order to achieve the ultimate estimation limit. One may then question: how can we find the optimal measurement? What kind of non-Gaussian measurement is required? We answer this question below, finally proposing optimal measurement operators, which has been often non-trivial~\cite{Nolan17}.

Let us begin with rewriting the QFI of Eq.~\eqref{QFI} in a more compact form as
\begin{align}
H(\phi)=\text{Tr}[\hat{\rho}_\phi \hat{L}_\phi^2],
\label{HwithL}
\end{align}
where $\hat{L}_\phi$ is the so-called symmetric logarithmic derivative (SLD) operator, defined in a way that
\begin{align}
\frac{\partial \hat{\rho}}{\partial \phi}=\frac{1}{2}\left(\hat{L}_\phi\hat{\rho}_\phi+\hat{\rho}_\phi \hat{L}_\phi\right).
\end{align}
The second equality in Eq.~\eqref{QCRB}, to which Eq.~\eqref{HwithL} is substituted, holds when two conditions are satisfied:
\begin{align}
&\text{Im}\left[\text{Tr}\left(\hat{\rho}_\phi\hat{\Pi}_k\hat{L}_\phi\right)\right]=0, \label{c1} \\
&\frac{\sqrt{\hat{\Pi}_k}\sqrt{\hat{\rho}_\phi}}{\text{Tr}[\hat{\rho}_\phi\hat{\Pi}_k]}=\frac{\sqrt{\hat{\Pi}_k}\hat{L}_\phi\sqrt{\hat{\rho}_\phi}}{\text{Tr}[\hat{\rho}_\phi\hat{\Pi}_k\hat{L}_\phi]}, \label{c2}
\end{align}
where $\hat{\rho}_{\phi}$ is the parameter $\phi$-encoded probe state. These conditions can be satisfied if one constitutes a POVM measurement setup $\{\Pi_k\}$ by a set of projection operators over the eigenbasis of $\hat{L}_\phi$~\cite{Braunstein94, Paris09}, so that the ultimate error bound given by the QFI is achieved. Especially, for a full-rank state of $\hat{\rho}_{\phi}$, the SLD operator is unique, and the above condition is a necessary and sufficient condition, i.e., the optimal setup prepared by the projection onto the eigenbasis of the SLD is the only optimal measurement. However, when $\hat{\rho}_{\phi}$ is a non-full-rank state, the SLD operator is not unique and other forms of SLD operators exist to determine the respective optimal measurement setups~\cite{Braunstein94}.

The SLD operator for a quantum state $\hat{\rho}=\sum_n p_n\ket{\psi_n}\bra{\psi_n}$ with $\langle\psi_n\vert\psi_m\rangle=\delta_{n,m}$ can be written as~\cite{Braunstein94, Paris09, Jiang14},
\begin{align}
\hat{L}_\phi=2 \sum_{n,m}\frac{\bra{\psi_m}\partial_\phi\rho_\phi\ket{\psi_n}}{p_n+p_m}\ket{\psi_m}\bra{\psi_n}, \label{SLD}
\end{align}
where the summation is taken over $n,m$ for which $p_n+p_m\neq 0$. A single-mode Gaussian state of Eq.~\eqref{lossy_input} can be spectrally decomposed as
\begin{align}
\hat{\rho}&=\sum_{n=0}^\infty p_n\hat{D}(\alpha)\hat{S}(\xi)|n\rangle\langle n|\hat{S}^\dagger(\xi)\hat{D}^\dagger(\alpha), \label{spectral}
\end{align}
where $p_n=n_\text{th}^n/(1+n_\text{th})^{n+1}$. Here $\hat{D}(\alpha)\hat{S}(\xi)\vert n\rangle$ and $\hat{D}(\alpha)\hat{S}(\xi)\vert m\rangle$ are orthonormal to each other when $n\neq m$. After some algebra (see Supplementary Section I for the detail), we then find the SLD operator for an arbitrary single-mode Gaussian state, which can be written as
\begin{align}
\hat{L}_\phi
&={\cal A}\hat{R}(\phi)\hat{S}(2\xi)\hat{D}(\zeta)\hat{R}(-\theta_{\rm s}/2)(\hat{X}\hat{P}+\hat{P}\hat{X})\nonumber\\
&\quad\quad\quad\times\hat{R}^{\dagger}(-\theta_{\rm s}/2)\hat{D}^{\dagger}(\zeta)\hat{S}^\dagger(2\xi)\hat{R}^\dagger(\phi) + {\cal C}\mathbb{I},
\label{generalSLD}
\end{align}
where 
\begin{align*}
{\cal A}&=\frac{(2n_{\rm th}+1)\sinh 2r}{2n_{\rm th}^{2}+2n_{\rm th}+1},\\
\zeta&=\alpha\cosh2r + \alpha^{*}e^{i\theta_{\rm s}}\left(\sinh 2r+\frac{1}{{\cal A}(2n_{\rm th}+1)}\right),\\
{\cal C}&=\frac{2|\alpha|^2}{{\cal A} (2n_\text{th}+1)^2}\sin (2\theta_{\rm c}-\theta_{\rm s}).
\end{align*}
Since the second term of the SLD in Eq.~\eqref{generalSLD} only rescales the eigenvalues, it can be absorbed into the post-data processing by an optimally chosen estimator. The first term, on the other hand, plays a crucial role in determining the optimal measurement setup that saturates the quantum Cram\'er-Rao bound. This indicates that for displaced squeezed thermal states, the optimal measurement setup needs to be constructed necessarily over the eigenbasis of $\hat{X}\hat{P}+\hat{P}\hat{X}$~\cite{Bollini93}. Therefore, this result reveals that any single-mode Gaussian measurement cannot be the optimal detection scheme for displaced squeezed thermal states. 

In particular, for displaced thermal states, the SLD of Eq.~\eqref{generalSLD} can be simplified (see Supplementary Section II for the detail) to be 
\begin{align}
\hat{L}_\phi
%=\frac{2i}{2n_{\rm th}+1}\hat{R}(\phi)\hat{D}(\alpha)(\alpha\hat{a}^{\dagger}-\alpha^{*}\hat{a}^{\dagger})\hat{D}^\dagger(\alpha)\hat{R}^\dagger(\phi),
=\frac{2\sqrt{2}\abs{\alpha}}{2n_{\rm th}+1}\hat{X}_{\theta_{\rm c}-\phi-\frac{\pi}{2}}.
\label{DTS_SLD}
\end{align}
This is the only optimal measurement setup for achieving the ultimate bound when displaced thermal states are used, as found also in Ref.~\cite{Monras2013}. It is apparent that the POVMs constructed over the eigenbasis of $\hat{L}_{\phi}$ in Eq.~\eqref{DTS_SLD} performs the homodyne detection. On the other hand, for squeezed thermal states, the SLD of Eq.~\eqref{generalSLD} is simplified (see Supplementary Section III for the detail) to be
\begin{align}
\hat{L}_{\phi}=\frac{(2n_{\rm th}+1)\sinh 2r}{2n_{\rm th}^{2}+2n_{\rm th}+1} \left(\hat{X}_{\theta_{\rm s}/2-\phi}\hat{P}_{\theta_{\rm s}/2-\phi}+\hat{P}_{\theta_{\rm s}/2-\phi}\hat{X}_{\theta_{\rm s}/2-\phi}\right).
\label{STS_SLD}
\end{align}
This is the only optimal measurement setup for the case of the squeezed thermal state input, and cannot be realized by any single-mode Gaussian measurement. This form has not been discussed elsewhere.
The SLD of Eq.~\eqref{STS_SLD} is valid also when $n_{\rm th}= 0$, i.e., when the probe state is a pure state, but in this case other type of optimal measurement apart from the SLD can exist; the homodyne detection has been shown to be optimal, satisfying the conditions in Eqs.~\eqref{c1} and \eqref{c2} although it is irrelevant to the SLD of Eq.~\eqref{STS_SLD} (see Supplementary Section IV for the proof).

In general, the SLD operator for an arbitrary single-mode Gaussian state can always be written in the form of~\cite{Gao2014,serafini2017, nichols2018}
\begin{align}\label{sld}
\hat{L}_{\phi}=L_\phi^{(0)}+L_\phi^{(1){\rm T}}\hat{Q}+\hat{Q}^{\rm T} L_\phi^{(2)}\hat{Q},
\end{align}
where $\hat{Q}=(\hat{X},\hat{P})^{\rm T}$ is the quadrature vector, $L_\phi^{(0)}$ is a real constant, $L_\phi^{(1)}$ is a real $2$-dimensional vector, and $L_\phi^{(2)}$ is a $2\times2$ real symmetric matrix.

One can easily show that $L_\phi^{(2)}$ is a zero matrix for the SLD operator of Eq.~\eqref{DTS_SLD}, so that the $\hat{L}_{\phi}$ for displaced thermal states is proportional to a rotated quadrature operator. This means that a homodyne detection is the optimal measurement.
The SLD operator of Eq.~\eqref{generalSLD}, on the other hand, has non-zero $L_\phi^{(2)}$, i.e., Eq.~\eqref{sld} takes the form of
\begin{align}\label{xppx}
\hat{L}_\phi=L_\phi^{(0)}+K\hat{Q}'^{\rm T}
\begin{pmatrix}
0 & 1 \\
1 & 0
\end{pmatrix}
\hat{Q}',
\end{align}
where $\hat{Q}'$ is a quadrature operator vector transformed by symplectic matrices and $K$ is a real constant. Our result shows that in this case, a non-Gaussian measurement is necessary to implement the optimal measurement for displaced squeezed thermal states. Note that Eq.~\eqref{xppx} can be written as being proportional to $i(\hat{a}^{2}-\hat{a}^{\dagger 2})$, but cannot be diagonalized by the Bogoliubov transformation that projects into the form of the photon number operator in the diagonalized basis~\cite{Blaizot85}.

It is also worth comparing the SLDs in Eqs.~\eqref{generalSLD}, \eqref{DTS_SLD} and \eqref{STS_SLD} with the SLD that has been found for the estimation of loss parameter in Gaussian metrology~\cite{Monras07}. For the case of loss parameter estimation in Gaussian metrology, the SLD operator takes the form of $\hat{L}_\phi=L_\phi^{(0)}+K\hat{Q}'^{\rm T}\mathbb{I}_{2}\hat{Q}'$. In this case, the Bogoluibov transformation can be applied to diagonalize it to the form of the photon number operator in the diagonalized basis, i.e., the optimal measurement for the loss parameter estimation requires the capability of photon number counting, whereas the eigenbasis of $\hat{X}\hat{P}+\hat{P}\hat{X}$ constructs the optimal setup for the phase parameter estimation. This difference implies that the kind of optimal setup depends on the type of parameter being estimated.
%\begin{align}\label{loss_SLD}
%\hat{L}_\phi=L_\phi^{(0)}+\mathcal{K}\hat{Q}'^{\rm T}
%\begin{pmatrix}
%1 & 0 \\
%0 & 1
%\end{pmatrix}
%\hat{Q}'
%%=L_\phi^{(0)}+\mathcal{K}(\hat{X}'^2+\hat{P}'^2)
%=L_\phi^{(0)}+\mathcal{K}{\hat{a}^\dagger}{'} \hat{a}',
%\end{align}

%%%%%%%%%%%%%%%%%%%%%%%%
%%%%%%%%%%%%%%%%%%%%%%%%
\section{Discussion} \label{sec:con}
%%%%%%%%%%%%%%%%%%%%%%%%
%%%%%%%%%%%%%%%%%%%%%%%%
In this work we have investigated the optimality of Gaussian measurements for phase estimation in single-mode Gaussian metrology. We have found the optimal Gaussian measurements for all kinds of single-mode Gaussian probe states, and then compared their associated FIs with the QFI obtainable with an optimal POVM. We have shown that for the cases using displaced thermal probe states or squeezed vacuum probe states, the Gaussian measurement (i.e., the homodyne detection) offers the ultimate optimal measurement setup, while for the other kinds of single-mode Gaussian probe states, the ultimate error bounds can be achieved only by the non-Gaussian measurement on the basis of eigenstate of the product quadrature operator $\hat{X}\hat{P}+\hat{P}\hat{X}$. With an analysis for squeezed thermal state inputs (i.e., with zero first moments), we have also demonstrated the counterexample against the conclusion given in Ref.~\cite{Monras2013} that the homodyne detection is optimal among all Gaussian measurements when first moments are fixed. Such remarkable discrepancy arises because Gaussian measurements taken into account in Ref.~\cite{Monras2013} do not include truly all Gaussian measurements.
Although the optimized Gaussian measurements are not fully optimal, they provide nearly optimal measurement setups in the limits when $n_{\rm th}\rightarrow 0$ or $\infty$, or when $\abs{\tilde{\alpha}}\rightarrow 0$ or $\infty$. These nearly optimal setups may be much more favored in an experiment, where practical imperfections tend to nullify the difference between optimal and nearly optimal setups. 

The way the work is carried out can be applied to multi-mode Gaussian metrologies \cite{Safranek15}, where entanglement starts to play an important role in parameter estimation. One may also investigate the optimality of Gaussian measurements for other types of parameter estimation, such as loss parameter estimation or frequency estimation. It would also be worthwhile to make a more rigorous analysis that interprets the role of input thermal photons $n_{\rm th,in}$ in the case of a displaced thermal state and a squeezed thermal state in future work. Furthermore, an experimental scheme to implement projections onto the eigenbasis of the optimal observable $\hat{X}\hat{P}+\hat{P}\hat{X}$ that reaches the fundamental bound is yet unknown, so we leave this for future study.
We also expect the effect of a non-trivial Gaussian environment, called the squeezing environment (e.g. see Ref.~\cite{Zarzyna17}), or even structured environments (e.g., see Ref.~\cite{Bina18}) to be studied in the near future.

%%%%%%%%%%%%%%%%%%%%%%%%
%%%%%%%%%%%%%%%%%%%%%%%%
\section*{acknowledgments}
%%%%%%%%%%%%%%%%%%%%%%%%
%%%%%%%%%%%%%%%%%%%%%%%%
S.-Y.L. is supported by Basic Science Research Program through the National Research Foundation of Korea (NRF) funded by the Ministry of Education (No.\ 2018R1D1A1B07048633). C.O. and H.J. are supported by a National Research Foundation of Korea grant funded by the Korea government (MSIP) (No.\ 2010-0018295) and by the KIST Institutional Program (No.\ 2E27800-18-P043). H.N. is supported by an NPRP grant 8-751-1-157 from Qatar National Research Fund.

\begin{widetext}

\section*{APPENDIX}

\begin{appendix}

%%%%%%%%%%%%%%%%%%%%%%%%
%%%%%%%%%%%%%%%%%%%%%%%%
\section{SLD for displaced squeezed thermal states}\label{Appendix_DSTS}
%%%%%%%%%%%%%%%%%%%%%%%%
%%%%%%%%%%%%%%%%%%%%%%%%
Here we derive the SLD for displaced squeezed thermal states. 
Eq.~(41) in the main text can be rewritten as
\begin{align}\tag{S1}
\hat{L}_\phi=2\sum_{n, m}\frac{p_n-p_m}{p_n+p_m}\langle\psi_m|\partial_\phi\psi_n\rangle\vert\psi_m\rangle\langle\psi_n\vert, \label{SLD2}
\end{align}
where the summation is taken over $n, m$, for which $p_n+p_m\neq0$.
Using Eq.~(42) in the main text and $|\partial_\phi\psi_n\rangle=-i\hat{a}^\dagger\hat{a}|\psi_n\rangle$ for phase rotation, one obtains
\begin{align}
\langle\psi_m\vert\partial_\phi\psi_n\rangle
&=-i\bra{m}S^\dagger(\xi)D^\dagger(\alpha)\hat{a}^\dagger\hat{a}\hat{D}(\alpha)\hat{S}(\xi)\ket{n}\nonumber\\
%=-i\langle m|(\hat{a}^\dagger\cosh{r}-\hat{a} e^{-i\theta_{\rm s}}\sinh{r}+\alpha^*)(\hat{a}\cosh{r}-\hat{a}^\dagger e^{i\theta_{\rm s}} \sinh{r}+\alpha)|n\rangle  \nonumber \\
&=-i\left(-\delta_{m-2,n}\sqrt{m}\sqrt{m-1}e^{i\theta_{\rm s}}\sinh{r}\cosh{r}\right.\nonumber \\
&\quad\quad\quad+\delta_{m-1,n}\sqrt{m}\alpha\cosh{r}-\delta_{m,n-2}\sqrt{n}\sqrt{n-1}e^{-i\theta_{\rm s}}\sinh{r}\cosh{r} \nonumber \\
&\quad\quad\quad\left.-\delta_{m,n-1}\sqrt{n}\alpha e^{-i\theta_{\rm s}}\sinh{r}+\delta_{m,n-1}\sqrt{n}\alpha^*\cosh{r}-\delta_{m-1,n}\sqrt{m}\alpha^*e^{i\theta_{\rm s}}\sinh{r}\right), \tag{S2}\label{partialSLD}
\end{align}
where the term proportional to $\delta_{m,n}$ is omitted as it is irrelevant in the summation.
Substituting $p_n=n_\text{th}^n/(1+n_\text{th})^{n+1}$ and Eq.~\eqref{partialSLD} into Eq.~\eqref{SLD2}, it is then simplified to be 
\begin{align}\tag{S3}
\hat{L}_\phi=\hat{R}(\phi)\hat{D}(\alpha)\hat{S}(\xi)(\hat{L}_1+\hat{L}_2)\hat{S}^\dagger(\xi)\hat{D}^\dagger(\alpha)\hat{R}^\dagger(\phi),
\label{Loriginal}
\end{align}
where 
\begin{align*} 
\hat{L}_1&=\frac{2}{2n_{\rm th}+1} i\left(\alpha^* \cosh{r}-\alpha e^{-i\theta_{\rm s}}\sinh{r}\right)\hat{a}+{\rm h.c.},\\
\hat{L}_2&=\frac{(2n_{\rm th}+1)\sinh{2r}}{2n_{\rm th}^2+2n_{\rm th}+1} i \hat{a}^{\dagger2}e^{i\theta_{\rm s}} +{\rm h.c.}.
\end{align*}
For a given operator $\hat{O}$ defined as
\begin{align*}
\hat{O}
&={\cal A}\hat{S}(\xi)\hat{D}(\beta)\hat{R}(-\theta_{\rm s}/2)(\hat{X}\hat{P}+\hat{P}\hat{X})\hat{R}^{\dagger}(-\theta_{\rm s}/2)\hat{D}^{\dagger}(\beta)\hat{S}^{\dagger}(\xi)\\
&=i{\cal A}\left[\hat{a}^{\dagger 2} e^{i\theta_{\rm s}}-\hat{a}^{2}e^{-i\theta_{\rm s}}\right]
+2i{\cal A}\left[\left (\beta\cosh re^{-i\theta_{\rm s}} - \beta^{*}\sinh r\right)\hat{a} 
+\left(\beta\sinh r-\beta^{*}e^{i\theta_{\rm s}} \cosh r\right)\hat{a}^{\dagger}
\right]
+i{\cal A}(\beta^{* 2}e^{i\theta_{\rm s}}-\beta^{2}e^{-i\theta_{\rm s}}),
\end{align*}
where $\beta=\abs{\beta}e^{i\theta_{b}}$, and assuming that $r\neq0$, it can easily be shown that when 
\begin{align*}
{\cal A}&=\frac{(2n_{\rm th}+1)\sinh 2r}{2n_{\rm th}^2+2n_{\rm th}+1},\\
\abs{\beta}&=\frac{\abs{\alpha}}{{\cal A}(2n_{\rm th}+1)},\\
\theta_{b}&=\theta_{\rm s} - \theta_{\rm c},
\end{align*}
the operator $\hat{L}_{1}+\hat{L}_{2}$ is written as 
\begin{align}\tag{S4}
\hat{L}_{1}+\hat{L}_{2}=\hat{O}-C\hat{\openone},
\label{replacement}
\end{align} 
where $C=2{\cal A}\abs{\beta}^{2}\sin (\theta_{\rm s}-2\theta_{c})$. Substituting Eq.~\eqref{replacement} to Eq.~\eqref{Loriginal}, Eq.~(43) in the main text is finally obtained after a little algebra. 

%%%%%%%%%%%%%%%%%%%%%%%%
%%%%%%%%%%%%%%%%%%%%%%%%
\section{SLD for displaced thermal states}\label{Appendix_DTS}
%%%%%%%%%%%%%%%%%%%%%%%%
%%%%%%%%%%%%%%%%%%%%%%%%
The SLD for displaced thermal states can be written as
\begin{align*}
\hat{L}_{\phi}=\frac{2 i}{2n_{\rm th}+1}\hat{R}(\phi)\hat{D}(\alpha)\left(\alpha^{*} \hat{a}-\alpha \hat{a}^{\dagger}\right)\hat{D}^{\dagger}(\alpha)\hat{R}^{\dagger}(\phi).
\end{align*}
This can be further simplified to be
\begin{align*}
\hat{L}_{\phi}
&=\frac{2 i}{2n_{\rm th}+1}\abs{\alpha}\left(\hat{a}e^{i(\phi-\theta_{\rm c})}-\hat{a}^{\dagger}e^{-i(\phi-\theta_{\rm c})}\right)
%&=\frac{2\sqrt{2}\abs{\alpha}}{2n_{\rm th}+1)}\hat{R}(\theta_{\rm c}+\phi)\hat{P}\hat{R}^{\dagger}(\theta_{\rm c}+\phi)\\
=\frac{2\sqrt{2}\abs{\alpha}}{2n_{\rm th}+1}\hat{X}_{\theta_{\rm c}-\phi-\frac{\pi}{2}}.
\end{align*}

%%%%%%%%%%%%%%%%%%%%%%%%
%%%%%%%%%%%%%%%%%%%%%%%%
\section{SLD for squeezed thermal states}\label{Appendix_STS}
%%%%%%%%%%%%%%%%%%%%%%%%
%%%%%%%%%%%%%%%%%%%%%%%%
The SLD for squeezed thermal states can be written as
\begin{align*}
\hat{L}_{\phi}=\frac{i(2n_{\rm th}+1)\sinh 2r}{2n_{\rm th}^{2}+2n_{\rm th}+1}\hat{R}(\phi)\hat{S}(\xi)\left(\hat{a}^{\dagger 2}e^{i\theta_{\rm s}}-\hat{a}^{2}e^{-i\theta_{\rm s}}\right)\hat{S}^{\dagger}(\xi)\hat{R}^{\dagger}(\phi).
\end{align*}
This can be further simplified to be
\begin{align*}
\hat{L}_{\phi}
%&=\frac{i(2n_{\rm th}+1)\sinh(2r)}{2n_{\rm th}^{2}+2n_{\rm th}+1}\hat{R}(\phi)\hat{S}(\xi)\hat{R}(\theta_{\rm s}/2)\left(\hat{a}^{\dagger 2}-\hat{a}^{2}\right)\hat{R}(\theta_{\rm s}/2)\hat{S}^{\dagger}(\xi)\hat{R}^{\dagger}(\phi)\\
&=\frac{i(2n_{\rm th}+1)\sinh 2r}{2n_{\rm th}^{2}+2n_{\rm th}+1}\hat{R}(\phi-\theta_{\rm s}/2)\left(\hat{a}^{\dagger 2}-\hat{a}^{2}\right)\hat{R}^{\dagger}(\phi-\theta_{\rm s}/2)\\
%&=\frac{(2n_{\rm th}+1)\sinh(2r)}{2n_{\rm th}^{2}+2n_{\rm th}+1}\hat{R}(\phi+\theta_{\rm s}/2+\pi)\left(\hat{X}\hat{P}+\hat{P}\hat{X}\right)\hat{R}^{\dagger}(\phi+\theta_{\rm s}/2+\pi)\\
&=\frac{(2n_{\rm th}+1)\sinh 2r}{2n_{\rm th}^{2}+2n_{\rm th}+1} \left(\hat{X}_{\theta_{\rm s}/2-\phi}\hat{P}_{\theta_{\rm s}/2-\phi}+\hat{P}_{\theta_{\rm s}/2-\phi}\hat{X}_{\theta_{\rm s}/2-\phi}\right).
\end{align*}

\hfill
%%%%%%%%%%%%%%%%%%%%%%%%
%%%%%%%%%%%%%%%%%%%%%%%%
\section{Optimality of the homodyne detection}\label{Appendix_Homodyne}
%%%%%%%%%%%%%%%%%%%%%%%%
%%%%%%%%%%%%%%%%%%%%%%%%
Here, we prove that homodyne detection is optimal for squeezed vacuum states explicitly by showing that homodyne detection satisfies Eqs.~(39) and (40) in the main text. 
First, one can easily verify that Eq.~(40) in the main text is automatically satisfied if the input state is pure and the POVM measurement setup is composed of rank-one projectors.

Now, we show that $\text{Tr}(\hat{\rho}_\phi\hat{\Pi}_k\hat{L}_\phi)$ is real.
Squeezed vacuum states and the SLD operator for the states can be written as
\begin{align}
\hat{\rho}_\phi&=\hat{R}(\phi)\hat{S}(\xi)|0\rangle\langle0|\hat{S}^\dagger(\xi)\hat{R}^\dagger(\phi)\nonumber\\
&=\hat{S}(\xi e^{-2i\phi})|0\rangle\langle0|\hat{S}^\dagger(\xi e^{-2i\phi}) \nonumber \\
&=|\xi e^{-2i\phi}\rangle\langle \xi e^{-2i\phi}|, \nonumber \\
\hat{L}_\phi&=i2\sinh2r\hat{R}(\phi)\hat{S}(\xi)(\hat{a}^{\dagger2}e^{i\theta_{\rm s}}-\hat{a}^2e^{-i\theta_{\rm s}})\hat{S}^\dagger(\xi)\hat{R}^\dagger(\phi). \nonumber
\end{align}
If we assume a homodyne detection with local oscillator angle $\psi/2$, $\hat{\Pi}_k=|x_{\psi/2}\rangle\langle x_{\psi/2}|$, which corresponds to the Gaussian measurement with the parameter $se^{i\psi}$ in the limit of $s\rightarrow\infty$,
\begin{align}
\text{Tr}\big(\hat{\rho}_\phi\hat{\Pi}_x\hat{L}_\phi\big)
&=i2\sinh2r\text{Tr}\left[|\xi e^{-2i\phi}\rangle\langle \xi e^{-2i\phi}|x_{\psi/2}\rangle\langle x_{\psi/2}|\hat{R}(\phi)\hat{S}(\xi)(\hat{a}^{\dagger2}e^{i\theta_{\rm s}}-\hat{a}^2e^{-i\theta_{\rm s}})\hat{S}^\dagger(\xi)\hat{R}^\dagger(\phi)\right] \nonumber \\
%&=i2\sqrt{2}\sinh2r e^{i\theta_{\rm s}}\langle \xi e^{-2i\phi}|x_{\psi/2}\rangle\langle x_{\psi/2}|\hat{R}(\phi)\hat{S}(\xi)|2\rangle \nonumber \\
%&=i2\sqrt{2}\sinh2r e^{i(\theta_{\rm s}-2\phi-\psi)}\langle \xi e^{i(-2\phi-\psi)}|x\rangle\langle x|\hat{S}(\xi e^{i(-2\phi-\psi)})|2\rangle \nonumber \\
%&=i2\sqrt{2}\sinh2r e^{i\chi}\langle r e^{i\chi}|x\rangle\langle x|\hat{S}(r e^{i\chi})|2\rangle, \nonumber\\
&=i2\sqrt{2}\sinh2r e^{i\chi}\langle r e^{i\chi}|x\rangle\left(\cosh^{-\frac{5}{2}}{r}\langle x|\exp(-e^{i\chi}\tanh{r}\frac{\hat{a}^{\dagger2}}{2})|2\rangle+\frac{\sqrt{2}}{2}e^{-i\chi}\tanh{r}\langle x|\hat{S}(r e^{i\chi})|0\rangle\right)\nonumber,
\end{align}
where $|x_\theta\rangle=\hat{R}^\dagger(\theta)|x\rangle$, and $\chi=\theta_{\rm s}-2\phi-\psi$. Here,
\begin{align*}
&\langle x|\exp(-e^{i\chi}\tanh{r}\frac{\hat{a}^{\dagger2}}{2})|2\rangle
%&=\sum_{m=0}^\infty \langle x|m\rangle\langle m|\exp(-e^{i\chi}\tanh{r}\frac{\hat{a}^{\dagger2}}{2})|2\rangle \nonumber \\
%&=\frac{\sqrt{2}}{\pi^{1/4}}e^{-x^2/2}\sum_{m=0}^\infty \frac{(-1)^me^{im\chi}\tanh^mr}{2^{2m+2} m!}H_{2m+2}(x) \nonumber \\
%&=-\frac{\sqrt{2}}{\pi^{1/4}}e^{-x^2/2}\sum_{m=0}^\infty(e^{i\chi}\tanh{r})^m (m+1) L_{m+1}^{-\frac{1}{2}}(x^2) \nonumber\\
=\frac{\sqrt{2}}{\pi^{1/4}}e^{-x^2/2}\frac{e^{\frac{e^{i\chi}\tanh{r}}{1-e^{i\chi}\tanh{r}}x^2}}{2(1-e^{i\chi}\tanh{r})^{5/2}}(e^{i\chi}\tanh{r}-1+2x^2),
\end{align*}
where we have used the derivative of generating function of Laguerre polynomials, given as
\begin{align*}
\sum_{m=0}^\infty t^m L_m^{(\alpha)}=\frac{1}{(1-t)^{\alpha+1}}e^{-\frac{tx}{1-t}}.
\end{align*}
Finally, after setting $\cos\chi=\tanh2r$ and using
\begin{align*}
\langle x|re^{i\chi}\rangle=\frac{\exp\bigg[-\frac{x^2}{2}\frac{\cosh{r}+e^{i\chi}\sinh{r}}{\cosh{r}-e^{i\chi}\sinh{r}}\bigg]}{\pi^{1/4}\sqrt{\cosh{r}-e^{i\chi}\sinh{r}}},
\end{align*}
we obtain
\begin{align*}
\text{Tr}\big(\hat{\rho}_\phi\hat{\Pi}_k\hat{L}_\phi\big)&=\frac{\exp(-x^2\cosh{2r})(2x^2\cosh{2r}-1)\sqrt{\cosh{2r}}}{\sqrt{2\pi}},
\end{align*}
which is real. This proves that the homodyne detection with the appropriate local oscillator angle is optimal for squeezed vacuum states.

\end{appendix}
\end{widetext}

%%%%%%%%%%%%%%%%%%%%%%%%
%%%%%%%%%%%%%%%%%%%%%%%%

\end{document}